\documentclass[aps,prl,twocolumn,preprintnumbers,amsmath,amssymb,floatfix]{revtex4-2}
\usepackage{graphicx}
\graphicspath{{./figures_bwr/}}
\usepackage{dcolumn}
\usepackage{bm}
\usepackage{gensymb}
\usepackage{textcomp}
\usepackage{mathtools}
\usepackage{braket}
\usepackage{xcolor}
\newif\ifshowrevisions
\showrevisionsfalse

\usepackage[colorlinks,linkcolor=blue,anchorcolor=blue,citecolor=blue,urlcolor=blue]{hyperref}

\begin{document}
\raggedbottom

\title{Photon-Statistics Control of Strong-Field Rydberg-State Excitation}

\author{Tao Jiang $^{1}$}
\thanks{These authors contributed equally to this work.}
\author{Jinlei Liu $^{1,2,3,\ast}$}
\email{liujinlei@nudt.edu.cn}
\author{Haopu Dou$^{1}$}
\author{Lingyi Zhao$^{1}$}
\author{Peng Xiang$^{1}$}
\author{Liangjun Peng$^{1}$}
\author{Renyu Wang$^{1}$}
\author{Jing Zhao$^{1,2,3}$}
\author{Xiaowei Wang$^{1,2,3}$}
\author{Yue Lang$^{1,2,3}$}
\author{Zengxiu Zhao $^{1,2,3}$}
\email{zhaozengxiu@nudt.edu.cn}

\affiliation{$^{1}$Department of Physics, National University of Defense Technology, Changsha 410073, China}
\affiliation{$^{2}$Hunan Key Laboratory of Extreme Matter and Applications, National University of Defense Technology, Changsha 410073, China}
\affiliation{$^{3}$Hunan Research Center of the Basic Discipline for Physical States, National University of Defense Technology, Changsha 410073, People's Republic of China}
\date{\today}

\begin{abstract}
Photon statistics provides a new control dimension for strong-field bound-state formation. 
Using a quantum-optical phase-space formulation coupled to time-dependent Schr\"odinger-equation calculations, we show that field fluctuations enhance multiphoton Rydberg excitation by broadening AC-Stark-shifted resonances, but suppress tunneling-mediated recapture even as ionization increases. A quantum-statistical recapture model attributes the latter to ensemble-induced dispersion of relative path phases. Photon statistics also reshapes Rydberg-manifold parity, linking field statistics to recapture coherence and final-state symmetry.
\end{abstract}

\maketitle

Strong-field electron dynamics is conventionally understood in terms of a prescribed laser waveform that launches and drives a quantum wave packet. 
This deterministic-field picture underlies the three-step model and the strong-field approximation for tunneling, recollision, and high-order harmonic generation~\cite{corkum_plasma_1993,lewenstein_theory_1994,krausz_attosecond_2009,milosevic_above-threshold_2006}. 
When the driving field itself possesses nontrivial photon statistics, however, the prescribed-waveform picture is no longer sufficient. The field fluctuations redistribute the weights of strong-field pathways and modify their interference after statistical averaging. 
Bright squeezed states are particularly suited to this regime because they combine macroscopic field energies with strongly non-Poissonian photon statistics and anisotropic quadrature fluctuations~\cite{walls_squeezed_1983,slusher_observation_1985,kolobov_spatial_1999,gerry_introductory_2023,gorlach2020quantum,cruz2024quantum,stammer2025colloquium}, allowing the electronic response to depart from that produced by a single deterministic waveform.

Recent advances in bright squeezed vacuum (BSV) have made this regime experimentally accessible. 
BSV combines macroscopic optical intensities with large photon-number fluctuations and superbunching~\cite{sh2012superbunched,perez2014bright,manceau_indefinite-mean_2019,sharapova2020properties}. Nonclassical light has now been used to drive high-order harmonic generation~\cite{rasputnyi2024high}, photon-bunched harmonic emission~\cite{lemieux2025photon}, strong-field photoemission from metal nanotips~\cite{heimerl2025quantum}, and BSV-driven atomic tunneling ionization and photoelectron statistics~\cite{jiang2026quantumboosted,liu2026strongfield}. Together with theoretical studies of quantum-light-driven high-order harmonic generation, above-threshold ionization, strong-field electron trajectories, and correlated ionization~\cite{gorlach2023high,even2023photon,fang2023strong,wang2023high,lyu2025effect,liu2025atomic}, these developments demonstrate that photon statistics can substantially modify continuum strong-field dynamics. Its role in bound-state formation, where returning electronic wave packets must be coherently recaptured rather than escape to the continuum, remains largely unexplored.

Rydberg-state excitation (RSE) provides a sensitive bound-state observable of photon-statistics-controlled strong-field dynamics,
because its yield reflects both population transfer and electron recapture. Under coherent driving, RSE arises through two distinct mechanisms. In the multiphoton regime, channel closings and AC-Stark-shifted resonances produce pronounced intensity-dependent structures in the Rydberg yield~\cite{li2014fine,chetty2020observation, 	yi2026intensity}. In the tunneling regime, frustrated tunneling ionization (FTI) describes bound-state formation through electron release, laser-driven excursion, and subsequent Coulomb-assisted recapture~\cite{zhang2014generation,hu2019quantum,chetty2022carrier}. The resulting Rydberg population retains information about the phase, energy, and angular-momentum content of the returning electron wave packet, making RSE particularly sensitive to
recapture-path interference. A central question is therefore how photon statistics modifies these two bound-state formation pathways under intense nonclassical driving, and whether electron release and bound-state formation can be controlled separately.

In this Letter, we establish photon statistics as a control parameter for strong-field Rydberg-state excitation. Using a quantum-optical phase-space formulation coupled to three-dimensional time-dependent Schr\"odinger-equation (TDSE) calculations, together with a quantum-statistical recapture (QSR) model, we investigate a hydrogen atom driven by coherent, BSV, and phase-squeezed coherent (PSC) fields. We find that field fluctuations enhance multiphoton RSE by broadening access to AC-Stark-shifted resonances, whereas tunneling-mediated RSE is suppressed even as ionization increases. The QSR model attributes this separation to the dispersion of relative phases between recapture pathways across the field ensemble, while the Rydberg-manifold parity provides a state-resolved signature of the modified recapture dynamics. These results establish
a bound-state avenue for photon-statistics control of strong-field electron dynamics and show that field statistics can modify bound-state formation separately from electron release.
\begin{figure}
	\centering
	\includegraphics[width=\columnwidth]{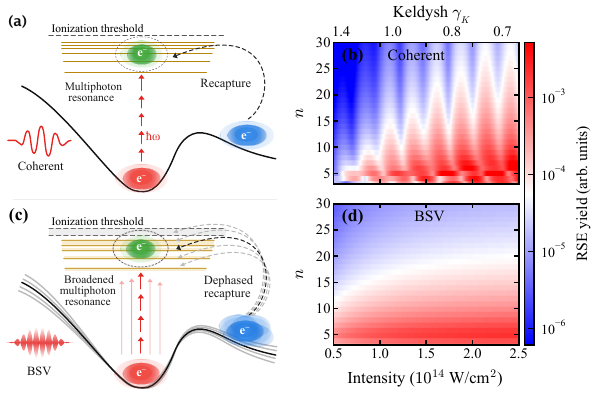}
	\caption{Photon-statistics modification of Rydberg-state excitation (RSE). (a),~(c) Schematic illustrations of RSE driven by coherent and bright-squeezed-vacuum (BSV) states, respectively. Rydberg states are populated either through multiphoton excitation into AC-Stark-shifted resonances below the ionization threshold or through tunneling release followed by electron recapture. A coherent field provides a deterministic waveform, leading to well-defined 	multiphoton resonance (red) conditions and phase-correlated recapture (black) pathways. In contrast, BSV represents a broad ensemble of 	coherent-field amplitudes, which broadens access to multiphoton resonances and reduces the contrast of recapture-path interference. (b),~(d) Principal-quantum-number-resolved RSE yields as functions of the mean peak intensity for coherent and BSV states, respectively. Calculations were performed for $\langle I\rangle=0.5$--$2.5\times10^{14}~\mathrm{W/cm^2}$ at a wavelength of $\lambda=800~\mathrm{nm}$.}
	\label{fig:1}
\end{figure}

The interaction of an intense quantum light field with a hydrogen atom is described by the joint atom--field density operator $\rho(t)$, which evolves according to
\begin{equation}
	i\hbar\frac{\partial\rho(t)}{\partial t}=
	\left[\hat{H}_{A}-\mathbf{r}\cdot\hat{\mathbf{E}}+\hat{H}_{F},\,\rho(t)\right].
	\label{eq1}
\end{equation}
Here, $\hat{H}_{A}=\mathbf{p}^{2}/2+V(\mathbf{r})$ is the atomic Hamiltonian with Coulomb potential $V(\mathbf{r})$, and
\begin{equation}
	\hat{H}_{F}=\sum_{\mathbf{k}}\hbar\omega_{\mathbf{k}}\left(\hat{a}_{\mathbf{k}}^{\dagger}\hat{a}_{\mathbf{k}}+\frac{1}{2}\right)
\end{equation}
is the free-field Hamiltonian. Within the electric-dipole approximation, the field operator is
\begin{equation}
	\hat{\mathbf{E}}=i\sum_{\mathbf{k}}\sqrt{\frac{\hbar\omega_{\mathbf{k}}}{2\epsilon_0 \mathcal V}}
	\,\mathbf{e}_{\mathbf{k}}\left(\hat{a}_{\mathbf{k}}-\hat{a}_{\mathbf{k}}^{\dagger}\right),
\end{equation}
where $\mathcal V$ is the field-normalization volume and $\mathbf{e}_{\mathbf{k}}$ is the polarization vector. We assume an initially factorized atom-field state $\rho(0)=\rho_{A}(0)\otimes\rho_{F}(0)$ and express the initial field state in the generalized $P$ representation \cite{scully1997quantum},
\begin{equation}
	\rho_{F}(0)=\int d^{2}\alpha d^{2}\beta\,P(\alpha,\beta^{*})
	\frac{|\alpha\rangle\langle\beta|}{\langle\beta|\alpha\rangle}.
\end{equation}

For the macroscopic photon numbers considered here, we evaluate
electronic population observables using a diagonal coherent-state
ensemble weighted by the Husimi distribution. This phase-space
formulation retains the photon statistics of the incident field through
$Q(\alpha)$: 
\begin{equation}
	\rho_{\mathrm{e}}(t)\simeq\int d^{2}\alpha\,Q(\alpha)
	|\phi_{\alpha}(t)\rangle\langle\phi_{\alpha}(t)|.
	\label{eq2}
\end{equation}
The electronic wave function $|\phi_{\alpha}(t)\rangle$ evolves under the coherent field $\mathbf{E}_{\alpha}(t)=\langle\alpha|\hat{\mathbf{E}}|\alpha\rangle$ according to
\begin{equation}
	i\hbar \frac{\partial |\phi_\alpha(t)\rangle}{\partial t}
	=
	\left[\hat H_A-\mathbf{r}\cdot \mathbf{E}_\alpha(t)\right]|\phi_\alpha(t)\rangle.
\end{equation}
Thus, the quantum statistics of the driving field enters through the ensemble of coherent-field realizations, whereas the electron dynamics
for each realization remains coherent \cite{gorlach2023high,even2023photon,wang2023high,fang2023strong,lyu2025effect,liu2025atomic,hillery1984distribution}. The finite field amplitude is $E_{\alpha}=2\xi^{(1)}\alpha$, with $\xi^{(1)}\propto \mathcal V^{-1/2}$. In the macroscopic-field limit $\mathcal V\rightarrow\infty$ with $E_{\alpha}$ fixed, distinct finite field amplitudes become effectively orthogonal after tracing over the incident mode, yielding the diagonal Husimi-weighted ensemble in Eq.~(\ref{eq2}) \cite{wang2025high}. 

The excitation probability into a specific field-free Rydberg state is given by
\begin{equation}
	Y_{nlm} = \int d^{2}\alpha\,Q(\alpha)\,p_{nlm}^{\alpha},
	\label{eq3}
\end{equation}
where $p_{nlm}^{\alpha}=\left|\langle \phi_{nlm} \mid \phi_{\alpha}(t_f)\rangle \right|^{2}$, $|\phi_{\alpha}(t_f)\rangle$ is the final electronic wave function, and $|\phi_{nlm}\rangle$ is a field-free Rydberg eigenstate. The
shell-resolved population is $Y_n=\sum_{l,m}Y_{nlm}$, and the total RSE yield reported below is summed over $3\le n\le 30$. 

For a displaced squeezed state $|\gamma,r\rangle$ with squeezing parameter $r$ and displacement $\gamma=\gamma_x+i\gamma_y$, the Husimi distribution is \cite{scully1997quantum,kim1989properties,even2023photon}
\begin{equation}
	Q(\alpha) = \frac{1}{\pi \cosh r}
	\exp \left[ -\frac{2(\alpha_y - \gamma_y)^2}{1 + e^{2r}}
	-\frac{2(\alpha_x - \gamma_x)^2}{1 + e^{-2r}} \right].
	\label{eq4}
\end{equation}
The coherent amplitude $\alpha=\alpha_{x}+i\alpha_{y}$ corresponds to the electric field $E_{\alpha}(t)=2\xi^{(1)}[\alpha_{x}\sin(\omega t)+\alpha_{y}\cos(\omega t)]$, with $\xi^{(1)}=\sqrt{\hbar\omega/(2\epsilon_0 \mathcal V)}$. For BSV, $\gamma=0$ and the mean intensity arises entirely from the squeezed-vacuum contribution, $I_{\mathrm{vac}}=c\hbar\omega \sinh^{2}(r)/\mathcal V = \epsilon_{0}c|E_{\mathrm{vac}}|^{2}/2$. For a phase-squeezed coherent (PSC) state $|\gamma,r\rangle$, the total intensity contains both coherent and squeezed-vacuum contributions, $I=\frac{\epsilon_{0}c}{2}(|E_{\gamma}|^{2}+|E_{\mathrm{vac}}|^{2})$ \cite{even2023photon}. Their relative weight is characterized by the dimensionless relative fluctuation strength $\chi=|E_{\mathrm{vac}}|^{2}/|E_{\gamma}|^{2}$, equivalently the squeezed-to-coherent intensity ratio. The coherent-field limit corresponds to $\chi\rightarrow0$, whereas the BSV limit is approached as $\chi\rightarrow\infty$.

For each coherent amplitude $\alpha$, we solve the three-dimensional TDSE \cite{tong1997theoretical,li2014fine,zhang2014generation} using a spherical-harmonic expansion up to $L_{\max}=50$ and a radial box extending to $r_{\max}=3000~\mathrm{a.u.}$ The initial state and field-free Rydberg eigenstates are obtained by diagonalizing $\hat H_A$ in the same numerical representation. Convergence of the Rydberg populations for $n\leq30$ was verified with respect to $L_{\max}$ and $r_{\max}$.

Figure~\ref{fig:1} illustrates the distinct RSE dynamics under coherent and BSV driving. The excitation dynamics span two regimes, including
a multiphoton-resonant regime and a  frustrated-tunneling-ionization
(FTI) regime~\cite{chetty2022carrier}, broadly associated with Keldysh parameters $\gamma_K>1$ and $\gamma_K<1$, respectively \cite{keldysh1965ionization,bing2006coulomb,nubbemeyer2008strong}. 
For coherent driving [Fig.~\ref{fig:1}(a)], the prescribed waveform fixes the field amplitude and phase. In the multiphoton regime, maxima in the Rydberg yield occur near channel-closing conditions, where the AC-Stark-shifted ionization threshold becomes resonant with an integer number of absorbed photons \cite{li2014fine,liu2021electron,yi2026intensity}. 
The resulting resonances of high-$n$ states recur with an intensity spacing determined by $\Delta U_p=\hbar\omega$, corresponding to approximately $26~\mathrm{TW/cm^2}$ at $800~\mathrm{nm}$. 
In the higher-intensity regime, RSE is dominated by FTI and is more naturally described in the time domain \cite{zimmermann_unified_2017}. 
The oscillatory structures in Fig.~\ref{fig:1}(b) then arise from interference among electron wave packets released in different half cycles and recaptured into the same Rydberg state \cite{xu2020observation,hu2019quantum}.

BSV modifies both regimes through its broad quantum-statistical distribution of coherent-field amplitudes. In the multiphoton regime [Fig.~\ref{fig:1}(c)], this distribution broadens the AC-Stark-shifted resonance conditions, so that the ensemble samples a wider range of near-resonant components at a fixed mean intensity. As a result, the sharp channel-closing structures produced by a
deterministic waveform are smoothed in the $n$-resolved population. In the FTI regime, different field realizations modify the relative
weights and accumulated phases of ionization-return pathways. Since the Rydberg population is averaged over these realizations, the recapture-path interference contrast is reduced without requiring
loss of coherence within any individual coherent-field component. This produces the smoother principal-quantum-number distributions
shown in Fig.~\ref{fig:1}(d). The effect is analogous to the fluctuation-induced washing out of above-threshold-ionization fringes predicted for quantum-light driving~\cite{lyu2025effect,liu2025atomic},
but here it is encoded in bound-state Rydberg populations.

\begin{figure}[!htbp]
	\centering
	\includegraphics[width=\columnwidth]{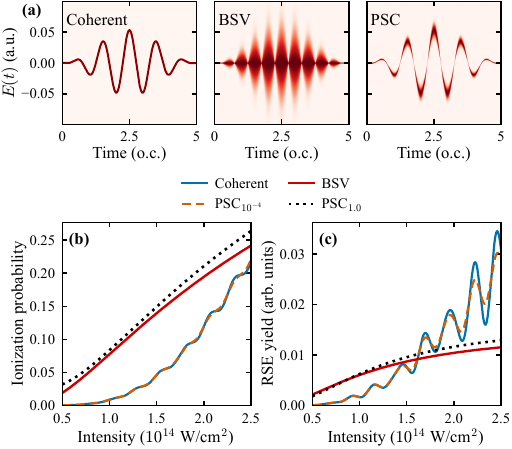}
	\caption{Regime-dependent response of ionization and Rydberg-state excitation 		to photon statistics. (a) Representative electric-field waveforms for coherent, BSV, and phase-squeezed coherent (PSC) states at the same mean peak intensity. (b) Ionization probabilities and (c) total Rydberg-state-excitation (RSE) yields, summed over $3\leq n\leq30$, as functions of the mean peak intensity. Results are shown for coherent and BSV states, together with PSC states at relative fluctuation strengths $\chi=10^{-4}$ and $1$. Field fluctuations enhance ionization over the displayed range, while their effect on RSE is regime dependent. RSE is enhanced in the multiphoton-dominated regime but suppressed in the FTI-dominated regime.}
	\label{fig:2}
\end{figure}

Figure~\ref{fig:2} quantifies how photon statistics changes electron release and bound-state formation. Figure~\ref{fig:2}(a) shows representative coherent-field realizations sampled from the
corresponding Husimi distributions. Coherent driving gives a single deterministic waveform, BSV samples zero-displacement amplitude fluctuations, and PSC provides an intermediate case in which a
coherent displacement is accompanied by amplitude fluctuations. As shown in Fig.~\ref{fig:2}(b), BSV enhances ionization throughout
the whole intensity range. In the multiphoton regime, this is consistent with the enhancement of higher-order absorption by photon bunching. For an $N$-photon process in the perturbative limit, the transition rate scales as
$W_N\propto g^{(N)}\langle I\rangle^N$, where
\begin{equation}
	g^{(N)}
	=
	\frac{\langle(\hat a^\dagger)^N\hat a^N\rangle}
	{\langle\hat a^\dagger\hat a\rangle^N}
\end{equation}
is the normalized $N$th-order field correlation function. Photon bunching therefore enhances multiphoton absorption for BSV, for which $g^{(N)}>1$, relative to coherent light, for which $g^{(N)}=1$ \cite{mollow1968two,agarwal1970field,spasibko_multiphoton_2017}.

The corresponding RSE response in Fig.~\ref{fig:2}(c) is strongly regime dependent. 
In the lower-intensity interval approaching the multiphoton-to-tunneling crossover, BSV enhances the total RSE yield. At fixed mean intensity, the BSV Husimi distribution spans a broad range of coherent-component amplitudes, increasing the
statistical weight of realizations that satisfy AC-Stark-shifted resonance conditions. The ensemble-averaged yield is therefore enhanced while the sharp intensity-dependent resonance structures are smoothed.

In the higher-intensity range,
$\langle I\rangle\gtrsim1.5\times10^{14}~{\rm W/cm^2}$, Fig.~\ref{fig:2}(c) shows the opposite behavior, where FTI-mediated RSE is suppressed even though ionization is enhanced. Electron release and
subsequent bound-state formation therefore respond differently to the same photon statistics. After tunneling, an electron can be driven back toward the ion and recaptured into a Rydberg orbital. Wave
packets released in different half cycles can reach the same final state and interfere, with their relative phases shaping the recapture yield~\cite{hu2019quantum,chetty2022carrier}. The decrease of the RSE yield in this regime therefore indicates a modification of recapture-path interference, which we analyze below using the
weak-fluctuation PSC case.

\begin{figure}[!t]
\centering
\includegraphics[width=\columnwidth]{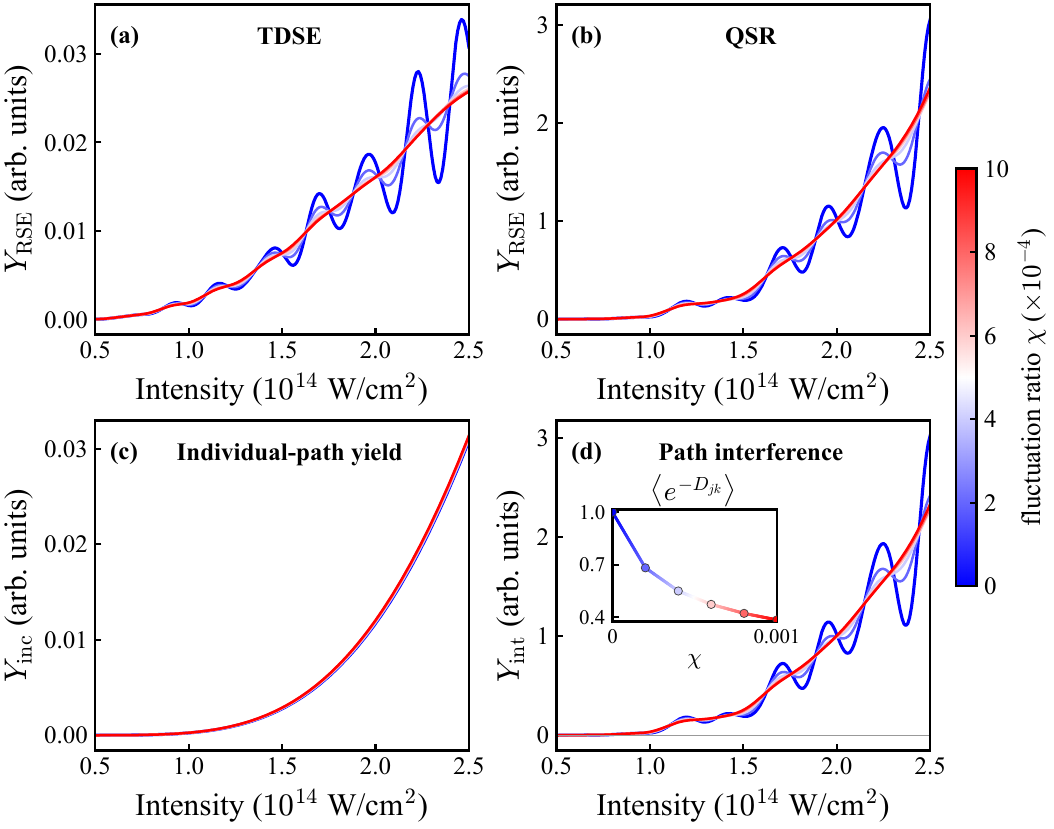}
\caption{Rydberg excitation and recapture-path interference under PSC states, with the relative fluctuation strength $\chi=0$--$10^{-3}$. RSE yields as functions of the mean peak intensity calculated using (a) TDSE and (b) a quantum-statistical recapture (QSR) model. (c) Incoherent individual-path contribution $Y_{\rm inc}$ and (d) interference contribution $Y_{\rm int}$ in the QSR model, as defined in Eq.~\eqref{eq:qsr_population} and summed over final states. Color encodes $\chi$. The inset shows the recapture-path pair coherence factor $\langle e^{-D_{jk}}\rangle$ at fixed total mean intensity $I=2\times10^{14}~\mathrm{W/cm^2}$. Increasing $\chi$ leaves the individual-path contribution nearly unchanged but progressively reduces the interference contrast, indicating that PSC fluctuations mainly suppress RSE through ensemble averaging of relative recapture phases.}
\label{fig:3}
\end{figure}

To identify the mechanism of the high-intensity suppression, we use a quantum-statistical recapture (QSR) model in the weak-fluctuation PSC regime. For a given coherent-field component $\alpha$, each
release-recapture path contributes an amplitude $b_j(\alpha)=a_j(\alpha)e^{-iS_j(\alpha)}$, where $a_j$ contains the tunneling-release and recapture weights and $S_j$ is the semiclassical phase acquired by the electron along its release-recapture trajectory in the laser field \cite{zhao2024twin}. For a given Rydberg state, the population is
\begin{equation}
\begin{aligned}
Y^{\rm QSR}&=\int d^2\alpha\,Q(\alpha)\left|\sum_j b_j(\alpha)\right|^2
=Y_{\rm inc}+Y_{\rm int},\\
Y_{\rm inc}&=\sum_j\int d^2\alpha\,Q(\alpha)|a_j(\alpha)|^2,\\
Y_{\rm int}&=2\operatorname{Re}\sum_{j<k}\int d^2\alpha\,Q(\alpha)a_j(\alpha)a_k^*(\alpha)e^{-i\Delta S_{jk}(\alpha)}.
\end{aligned}
\label{eq:qsr_population}
\end{equation}
Here $\Delta S_{jk}(\alpha)=S_j(\alpha)-S_k(\alpha)$ is the relative action of two recapture paths. The electron
amplitudes interfere coherently within each field realization, and the population is averaged over the incident-field statistics.

Figures~\ref{fig:3}(a) and \ref{fig:3}(b) show that the QSR model reproduces the fluctuation-induced smoothing and suppression of the RSE yield found in the TDSE calculations. The decomposition in Figs.~\ref{fig:3}(c) and \ref{fig:3}(d) identifies the origin of this suppression. The individual-path contribution $Y_{\rm inc}$ increases slightly as $\chi$ grows. By contrast, the interference contribution $Y_{\rm int}$ exhibits strong intensity-dependent fringes for coherent driving, but these fringes lose contrast under PSC fluctuations. Thus the high-intensity suppression of RSE is not caused primarily by a reduction of the single-path recapture probability. It results from the statistical averaging of phase-sensitive interference between different release-recapture paths.

For the PSC state $|\gamma,r\rangle$, the field amplitude fluctuates about its displacement value, $E=E_\gamma+\xi$. In the macroscopic-field limit, the Husimi marginal gives Gaussian deviations $Q(\xi)=\exp[-\xi^2/(2|E_{\rm vac}|^2)]/(\sqrt{2\pi}|E_{\rm vac}|)$ with variance $|E_{\rm vac}|^2=\chi|E_\gamma|^2$. For weak relative fluctuations, we keep the path weights, the release and recapture times at their values in the displacement field $E_\gamma$ and expand $S_j(\xi)\simeq S_j^\gamma+\lambda_j\xi$, where $S_j^\gamma$ is the action in the PSC displacement field and $\lambda_j=\partial S_j/\partial\xi$ is evaluated at $E_\gamma$. The relative action (see Supplemental Material \cite{supplemental_material} for the action derivative) then becomes
\begin{equation}
	\Delta S_{jk}(\xi)
	\simeq
	\Delta S_{jk}^\gamma+(\lambda_j-\lambda_k)\xi ,
\end{equation}
and averaging the corresponding phase factor gives
\begin{equation}
\begin{aligned}
	\int d\xi\,Q(\xi)e^{-i\Delta S_{jk}(\xi)}
	&\simeq
	e^{-i\Delta S_{jk}^\gamma}e^{-D_{jk}},
	\\
	D_{jk}
	&=
	\frac{|E_{\rm vac}|^2}{2}
	(\lambda_j-\lambda_k)^2 .
\end{aligned}
\label{eq:qsr_dephasing}
\end{equation}

The factor $e^{-D_{jk}}$ is a pair-coherence factor for two recapture paths. It depends only on the differential field sensitivity $\lambda_j-\lambda_k$. If two paths respond identically to the field fluctuation, their relative phase is preserved, whereas paths with
different sensitivities acquire a distribution of relative phases across the ensemble. Since $D_{jj}=0$, the statistical phase averaging acts on path interference rather than on an
individual recapture path. The inset of Fig.~\ref{fig:3} shows that the averaged factor $\langle e^{-D_{jk}}\rangle$ decreases with
$\chi$, consistent with the loss of contrast in $Y_{\rm int}$ and the smoothing of the total RSE yield.

Beyond modifying the total excitation yield and recapture interference, photon statistics also affects the spatial symmetry of the final Rydberg manifold. To quantify this effect, we define the
normalized $l$-resolved population
\begin{equation}
	P_l(I)
	=
	\frac{
		\sum_{n,m} p_{nlm}(I)
	}{
		\sum_{n,l,m} p_{nlm}(I)
	},
\end{equation}
and the Rydberg-manifold parity
\begin{equation}
	\mathcal P_R(I)
	=
	\frac{P_{\rm even}(I)-P_{\rm odd}(I)}
	{P_{\rm even}(I)+P_{\rm odd}(I)} ,
\end{equation}
where
$P_{\rm even(odd)}(I)=\sum_{l\in{\rm even(odd)}}P_l(I)$.
Thus, $\mathcal P_R$ measures the parity polarization of the excited
Rydberg population: $\mathcal P_R>0$ and $\mathcal P_R<0$ correspond
to predominantly even- and odd-parity excitation, respectively, while
$\mathcal P_R\simeq0$ indicates a parity-mixed Rydberg manifold.

\begin{figure}[!t]
	\centering
	\includegraphics[width=\columnwidth]{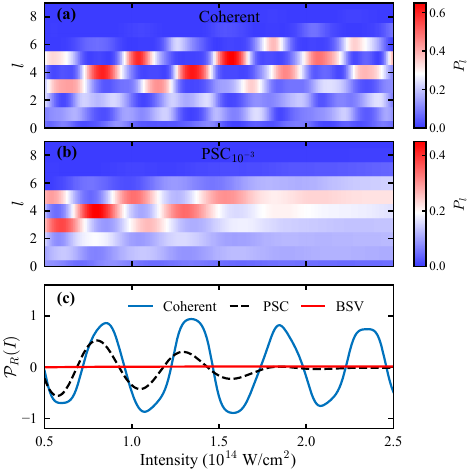}
	\caption{Photon-statistical imprint on the parity composition of the final Rydberg manifold. (a), (b) Normalized orbital-angular-momentum distributions of the Rydberg population for coherent driving and PSC driving with $\chi=10^{-3}$, respectively. 	(c) Rydberg-manifold parity $\mathcal{P}_{R}(I)$ as a function of the mean peak intensity for coherent, PSC, and BSV driving. Positive (negative) values of $\mathcal{P}_{R}$ indicate predominantly even- (odd-) parity excitation, whereas  $\mathcal{P}_{R}\simeq0$ corresponds to a parity-mixed manifold. Increasing field fluctuations drive the strongly parity-selective coherent-field response toward a parity-mixed final-state distribution.
	}
	\label{fig:4}
\end{figure}

Figure~\ref{fig:4} shows that photon statistics leaves a state-resolved imprint on the final Rydberg manifold. For coherent driving [Fig.~\ref{fig:4}(a)], the $l$-resolved population exhibits a pronounced even-odd alternation that changes with laser intensity. In the multiphoton regime, this parity selectivity reflects the
parity of the dominant photon-absorption channel
\cite{arbo2008sub,venzke2018angular,liu2024parity}. In the FTI regime, by contrast, the final-state parity is determined by the coherent superposition of release-recapture pathways leading to the
same Rydberg manifold, and is therefore sensitive to their relative phases and to the temporal symmetry of the few-cycle field \cite{hu2019quantum,chetty2022carrier,chetty2020observation}.

Field fluctuations qualitatively reshape this parity structure. For PSC driving with $\chi=10^{-3}$ [Fig.~\ref{fig:4}(b)], the even--odd
alternation remains visible at lower intensities but is strongly reduced in the FTI regime, where both parity sectors become populated. The same trend is summarized by $\mathcal P_R(I)$ in
Fig.~\ref{fig:4}(c), where the large parity polarization generated by coherent driving is reduced for PSC light and approaches a near-parity-mixed distribution for BSV. This evolution is consistent with the recapture-interference picture in Fig.~\ref{fig:3}. When field fluctuations broaden the distribution of relative phases between release-recapture pathways, they reduce not only the total interference contrast in the Rydberg yield but also the parity selectivity encoded in the final bound state.

In conclusion, we have established photon statistics as a control parameter for the bound-state branch of strong-field electron
dynamics. Combining a quantum-optical phase-space formulation with TDSE calculations and a QSR model, we showed that field fluctuations reshape Rydberg-state excitation in a regime-dependent manner. In the multiphoton regime, the broad distribution of coherent-field amplitudes enhances the statistical sampling of AC-Stark-shifted resonances and increases the Rydberg yield. In the FTI regime, however, electron release is enhanced while subsequent bound-state formation is suppressed. The QSR analysis attributes this separation to ensemble-induced dispersion of the relative phases
between release--recapture pathways, which reduces recapture-path interference after population averaging.

Photon statistics also leaves a state-resolved imprint on the final Rydberg manifold. The pronounced even-odd orbital-angular-momentum
selectivity generated by coherent driving evolves toward a parity-mixed distribution as the field fluctuations increase, demonstrating that photon statistics modifies not only the probability
of bound-state formation but also the discrete spatial symmetry of the final electronic state. These results identify Rydberg-state excitation as a bound-state probe of photon-statistics-controlled strong-field dynamics and open a route to controlling strong-field final-state distributions with nonclassical light.

\section*{Acknowledgements} \label{sec:acknowledgements}
This work was supported by the National Natural Science Foundation of China (Grant Nos.~12234020, 12274461, 12595340, and 12304306), the Hunan Provincial Major Basic Research Program (Grant No.~2025ZYJ001), and the Innovation Research Foundation of NUDT (Grant Nos.~202501-RCGC-ZZ-001, 26-ZZCX-JDZ-05, 25-ZZCX-DFXJS-01, and ZK23-04).


\begin{thebibliography}{54}%
	\makeatletter
	\providecommand \@ifxundefined [1]{%
		\@ifx{#1\undefined}
	}%
	\providecommand \@ifnum [1]{%
		\ifnum #1\expandafter \@firstoftwo
		\else \expandafter \@secondoftwo
		\fi
	}%
	\providecommand \@ifx [1]{%
		\ifx #1\expandafter \@firstoftwo
		\else \expandafter \@secondoftwo
		\fi
	}%
	\providecommand \natexlab [1]{#1}%
	\providecommand \enquote  [1]{``#1''}%
	\providecommand \bibnamefont  [1]{#1}%
	\providecommand \bibfnamefont [1]{#1}%
	\providecommand \citenamefont [1]{#1}%
	\providecommand \href@noop [0]{\@secondoftwo}%
	\providecommand \href [0]{\begingroup \@sanitize@url \@href}%
	\providecommand \@href[1]{\@@startlink{#1}\@@href}%
	\providecommand \@@href[1]{\endgroup#1\@@endlink}%
	\providecommand \@sanitize@url [0]{\catcode `\\12\catcode `\$12\catcode
		`\&12\catcode `\#12\catcode `\^12\catcode `\_12\catcode `\%12\relax}%
	\providecommand \@@startlink[1]{}%
	\providecommand \@@endlink[0]{}%
	\providecommand \url  [0]{\begingroup\@sanitize@url \@url }%
	\providecommand \@url [1]{\endgroup\@href {#1}{\urlprefix }}%
	\providecommand \urlprefix  [0]{URL }%
	\providecommand \Eprint [0]{\href }%
	\providecommand \doibase [0]{https://doi.org/}%
	\providecommand \selectlanguage [0]{\@gobble}%
	\providecommand \bibinfo  [0]{\@secondoftwo}%
	\providecommand \bibfield  [0]{\@secondoftwo}%
	\providecommand \translation [1]{[#1]}%
	\providecommand \BibitemOpen [0]{}%
	\providecommand \bibitemStop [0]{}%
	\providecommand \bibitemNoStop [0]{.\EOS\space}%
	\providecommand \EOS [0]{\spacefactor3000\relax}%
	\providecommand \BibitemShut  [1]{\csname bibitem#1\endcsname}%
	\let\auto@bib@innerbib\@empty
	\bibitem [{\citenamefont {Corkum}(1993)}]{corkum_plasma_1993}%
	\BibitemOpen
	\bibfield  {author} {\bibinfo {author} {\bibfnamefont {P.~B.}\ \bibnamefont
			{Corkum}},\ }\bibfield  {title} {\bibinfo {title} {Plasma perspective on
			strong field multiphoton ionization},\ }\href
	{https://doi.org/10.1103/PhysRevLett.71.1994} {\bibfield  {journal} {\bibinfo
			{journal} {Physical Review Letters}\ }\textbf {\bibinfo {volume} {71}},\
		\bibinfo {pages} {1994} (\bibinfo {year} {1993})}\BibitemShut {NoStop}%
	\bibitem [{\citenamefont {Lewenstein}\ \emph {et~al.}(1994)\citenamefont
		{Lewenstein}, \citenamefont {Balcou}, \citenamefont {Ivanov}, \citenamefont
		{L'Huillier},\ and\ \citenamefont {Corkum}}]{lewenstein_theory_1994}%
	\BibitemOpen
	\bibfield  {author} {\bibinfo {author} {\bibfnamefont {M.}~\bibnamefont
			{Lewenstein}}, \bibinfo {author} {\bibfnamefont {P.}~\bibnamefont {Balcou}},
		\bibinfo {author} {\bibfnamefont {M.~Y.}\ \bibnamefont {Ivanov}}, \bibinfo
		{author} {\bibfnamefont {A.}~\bibnamefont {L'Huillier}},\ and\ \bibinfo
		{author} {\bibfnamefont {P.~B.}\ \bibnamefont {Corkum}},\ }\bibfield  {title}
	{\bibinfo {title} {Theory of high-harmonic generation by low-frequency laser
			fields},\ }\href {https://doi.org/10.1103/PhysRevA.49.2117} {\bibfield
		{journal} {\bibinfo  {journal} {Physical Review A}\ }\textbf {\bibinfo
			{volume} {49}},\ \bibinfo {pages} {2117} (\bibinfo {year}
		{1994})}\BibitemShut {NoStop}%
	\bibitem [{\citenamefont {Krausz}\ and\ \citenamefont
		{Ivanov}(2009)}]{krausz_attosecond_2009}%
	\BibitemOpen
	\bibfield  {author} {\bibinfo {author} {\bibfnamefont {F.}~\bibnamefont
			{Krausz}}\ and\ \bibinfo {author} {\bibfnamefont {M.}~\bibnamefont
			{Ivanov}},\ }\bibfield  {title} {\bibinfo {title} {Attosecond physics},\
	}\href {https://doi.org/10.1103/RevModPhys.81.163} {\bibfield  {journal}
		{\bibinfo  {journal} {Reviews of Modern Physics}\ }\textbf {\bibinfo {volume}
			{81}},\ \bibinfo {pages} {163} (\bibinfo {year} {2009})}\BibitemShut
	{NoStop}%
	\bibitem [{\citenamefont {Milošević}\ \emph {et~al.}(2006)\citenamefont
		{Milošević}, \citenamefont {Paulus}, \citenamefont {Bauer},\ and\
		\citenamefont {Becker}}]{milosevic_above-threshold_2006}%
	\BibitemOpen
	\bibfield  {author} {\bibinfo {author} {\bibfnamefont {D.~B.}\ \bibnamefont
			{Milošević}}, \bibinfo {author} {\bibfnamefont {G.~G.}\ \bibnamefont
			{Paulus}}, \bibinfo {author} {\bibfnamefont {D.}~\bibnamefont {Bauer}},\ and\
		\bibinfo {author} {\bibfnamefont {W.}~\bibnamefont {Becker}},\ }\bibfield
	{title} {\bibinfo {title} {Above-threshold ionization by few-cycle pulses},\
	}\href {https://doi.org/10.1088/0953-4075/39/14/R01} {\bibfield  {journal}
		{\bibinfo  {journal} {Journal of Physics B: Atomic, Molecular and Optical
				Physics}\ }\textbf {\bibinfo {volume} {39}},\ \bibinfo {pages} {R203}
		(\bibinfo {year} {2006})}\BibitemShut {NoStop}%
	\bibitem [{\citenamefont {Walls}(1983)}]{walls_squeezed_1983}%
	\BibitemOpen
	\bibfield  {author} {\bibinfo {author} {\bibfnamefont {D.~F.}\ \bibnamefont
			{Walls}},\ }\bibfield  {title} {\bibinfo {title} {Squeezed states of light},\
	}\href {https://doi.org/10.1038/306141a0} {\bibfield  {journal} {\bibinfo
			{journal} {Nature}\ }\textbf {\bibinfo {volume} {306}},\ \bibinfo {pages}
		{141} (\bibinfo {year} {1983})}\BibitemShut {NoStop}%
	\bibitem [{\citenamefont {Slusher}\ \emph {et~al.}(1985)\citenamefont
		{Slusher}, \citenamefont {Hollberg}, \citenamefont {Yurke}, \citenamefont
		{Mertz},\ and\ \citenamefont {Valley}}]{slusher_observation_1985}%
	\BibitemOpen
	\bibfield  {author} {\bibinfo {author} {\bibfnamefont {R.~E.}\ \bibnamefont
			{Slusher}}, \bibinfo {author} {\bibfnamefont {L.~W.}\ \bibnamefont
			{Hollberg}}, \bibinfo {author} {\bibfnamefont {B.}~\bibnamefont {Yurke}},
		\bibinfo {author} {\bibfnamefont {J.~C.}\ \bibnamefont {Mertz}},\ and\
		\bibinfo {author} {\bibfnamefont {J.~F.}\ \bibnamefont {Valley}},\ }\bibfield
	{title} {\bibinfo {title} {Observation of {Squeezed} {States} {Generated} by
			{Four}-{Wave} {Mixing} in an {Optical} {Cavity}},\ }\href
	{https://doi.org/10.1103/PhysRevLett.55.2409} {\bibfield  {journal} {\bibinfo
			{journal} {Physical Review Letters}\ }\textbf {\bibinfo {volume} {55}},\
		\bibinfo {pages} {2409} (\bibinfo {year} {1985})}\BibitemShut {NoStop}%
	\bibitem [{\citenamefont {Kolobov}(1999)}]{kolobov_spatial_1999}%
	\BibitemOpen
	\bibfield  {author} {\bibinfo {author} {\bibfnamefont {M.~I.}\ \bibnamefont
			{Kolobov}},\ }\bibfield  {title} {\bibinfo {title} {The spatial behavior of
			nonclassical light},\ }\href {https://doi.org/10.1103/RevModPhys.71.1539}
	{\bibfield  {journal} {\bibinfo  {journal} {Reviews of Modern Physics}\
		}\textbf {\bibinfo {volume} {71}},\ \bibinfo {pages} {1539} (\bibinfo {year}
		{1999})}\BibitemShut {NoStop}%
	\bibitem [{\citenamefont {Gerry}\ and\ \citenamefont
		{Knight}(2023)}]{gerry_introductory_2023}%
	\BibitemOpen
	\bibfield  {author} {\bibinfo {author} {\bibfnamefont {C.}~\bibnamefont
			{Gerry}}\ and\ \bibinfo {author} {\bibfnamefont {P.}~\bibnamefont {Knight}},\
	}\href {https://books.google.co.jp/books?id=p-_kEAAAQBAJ} {\emph {\bibinfo
			{title} {Introductory {Quantum} {Optics}}}}\ (\bibinfo  {publisher}
	{Cambridge University Press},\ \bibinfo {year} {2023})\BibitemShut {NoStop}%
	\bibitem [{\citenamefont {Gorlach}\ \emph {et~al.}(2020)\citenamefont
		{Gorlach}, \citenamefont {Neufeld}, \citenamefont {Rivera}, \citenamefont
		{Cohen},\ and\ \citenamefont {Kaminer}}]{gorlach2020quantum}%
	\BibitemOpen
	\bibfield  {author} {\bibinfo {author} {\bibfnamefont {A.}~\bibnamefont
			{Gorlach}}, \bibinfo {author} {\bibfnamefont {O.}~\bibnamefont {Neufeld}},
		\bibinfo {author} {\bibfnamefont {N.}~\bibnamefont {Rivera}}, \bibinfo
		{author} {\bibfnamefont {O.}~\bibnamefont {Cohen}},\ and\ \bibinfo {author}
		{\bibfnamefont {I.}~\bibnamefont {Kaminer}},\ }\bibfield  {title} {\bibinfo
		{title} {The quantum-optical nature of high harmonic generation},\ }\href
	{https://www.nature.com/articles/s41467-020-18218-w} {\bibfield  {journal}
		{\bibinfo  {journal} {Nature Communications}\ }\textbf {\bibinfo {volume}
			{11}},\ \bibinfo {pages} {4598} (\bibinfo {year} {2020})}\BibitemShut
	{NoStop}%
	\bibitem [{\citenamefont {Cruz-Rodriguez}\ \emph {et~al.}(2024)\citenamefont
		{Cruz-Rodriguez}, \citenamefont {Dey}, \citenamefont {Freibert},\ and\
		\citenamefont {Stammer}}]{cruz2024quantum}%
	\BibitemOpen
	\bibfield  {author} {\bibinfo {author} {\bibfnamefont {L.}~\bibnamefont
			{Cruz-Rodriguez}}, \bibinfo {author} {\bibfnamefont {D.}~\bibnamefont {Dey}},
		\bibinfo {author} {\bibfnamefont {A.}~\bibnamefont {Freibert}},\ and\
		\bibinfo {author} {\bibfnamefont {P.}~\bibnamefont {Stammer}},\ }\bibfield
	{title} {\bibinfo {title} {Quantum phenomena in attosecond science},\
	}\href@noop {} {\bibfield  {journal} {\bibinfo  {journal} {Nature Reviews
				Physics}\ }\textbf {\bibinfo {volume} {6}},\ \bibinfo {pages} {691} (\bibinfo
		{year} {2024})}\BibitemShut {NoStop}%
	\bibitem [{\citenamefont {Stammer}\ \emph {et~al.}(2025)\citenamefont
		{Stammer}, \citenamefont {Rivera-Dean}, \citenamefont {Tzallas},
		\citenamefont {Ciappina},\ and\ \citenamefont
		{Lewenstein}}]{stammer2025colloquium}%
	\BibitemOpen
	\bibfield  {author} {\bibinfo {author} {\bibfnamefont {P.}~\bibnamefont
			{Stammer}}, \bibinfo {author} {\bibfnamefont {J.}~\bibnamefont
			{Rivera-Dean}}, \bibinfo {author} {\bibfnamefont {P.}~\bibnamefont
			{Tzallas}}, \bibinfo {author} {\bibfnamefont {M.~F.}\ \bibnamefont
			{Ciappina}},\ and\ \bibinfo {author} {\bibfnamefont {M.}~\bibnamefont
			{Lewenstein}},\ }\bibfield  {title} {\bibinfo {title} {Colloquium: Quantum
			optics of intense light--matter interaction},\ }\href@noop {} {\bibfield
		{journal} {\bibinfo  {journal} {arXiv preprint arXiv:2510.19045}\ } (\bibinfo
		{year} {2025})}\BibitemShut {NoStop}%
	\bibitem [{\citenamefont {Sh.~Iskhakov}\ \emph {et~al.}(2012)\citenamefont
		{Sh.~Iskhakov}, \citenamefont {P{\'e}rez}, \citenamefont {Yu.~Spasibko},
		\citenamefont {Chekhova},\ and\ \citenamefont {Leuchs}}]{sh2012superbunched}%
	\BibitemOpen
	\bibfield  {author} {\bibinfo {author} {\bibfnamefont {T.}~\bibnamefont
			{Sh.~Iskhakov}}, \bibinfo {author} {\bibfnamefont {A.~M.}\ \bibnamefont
			{P{\'e}rez}}, \bibinfo {author} {\bibfnamefont {K.}~\bibnamefont
			{Yu.~Spasibko}}, \bibinfo {author} {\bibfnamefont {M.~V.}\ \bibnamefont
			{Chekhova}},\ and\ \bibinfo {author} {\bibfnamefont {G.}~\bibnamefont
			{Leuchs}},\ }\bibfield  {title} {\bibinfo {title} {Superbunched bright
			squeezed vacuum state},\ }\href
	{https://opg.optica.org/abstract.cfm?URI=ol-37-11-1919} {\bibfield  {journal}
		{\bibinfo  {journal} {Optics Letters}\ }\textbf {\bibinfo {volume} {37}},\
		\bibinfo {pages} {1919} (\bibinfo {year} {2012})}\BibitemShut {NoStop}%
	\bibitem [{\citenamefont {P{\'e}rez}\ \emph {et~al.}(2014)\citenamefont
		{P{\'e}rez}, \citenamefont {Iskhakov}, \citenamefont {Sharapova},
		\citenamefont {Lemieux}, \citenamefont {Tikhonova}, \citenamefont
		{Chekhova},\ and\ \citenamefont {Leuchs}}]{perez2014bright}%
	\BibitemOpen
	\bibfield  {author} {\bibinfo {author} {\bibfnamefont {A.~M.}\ \bibnamefont
			{P{\'e}rez}}, \bibinfo {author} {\bibfnamefont {T.~{\relax Sh}.}\
			\bibnamefont {Iskhakov}}, \bibinfo {author} {\bibfnamefont {P.}~\bibnamefont
			{Sharapova}}, \bibinfo {author} {\bibfnamefont {S.}~\bibnamefont {Lemieux}},
		\bibinfo {author} {\bibfnamefont {O.~V.}\ \bibnamefont {Tikhonova}}, \bibinfo
		{author} {\bibfnamefont {M.~V.}\ \bibnamefont {Chekhova}},\ and\ \bibinfo
		{author} {\bibfnamefont {G.}~\bibnamefont {Leuchs}},\ }\bibfield  {title}
	{\bibinfo {title} {Bright squeezed-vacuum source with 11 spatial mode},\
	}\href {https://opg.optica.org/abstract.cfm?URI=ol-39-8-2403} {\bibfield
		{journal} {\bibinfo  {journal} {Optics Letters}\ }\textbf {\bibinfo {volume}
			{39}},\ \bibinfo {pages} {2403} (\bibinfo {year} {2014})}\BibitemShut
	{NoStop}%
	\bibitem [{\citenamefont {Manceau}\ \emph {et~al.}(2019)\citenamefont
		{Manceau}, \citenamefont {Spasibko}, \citenamefont {Leuchs}, \citenamefont
		{Filip},\ and\ \citenamefont {Chekhova}}]{manceau_indefinite-mean_2019}%
	\BibitemOpen
	\bibfield  {author} {\bibinfo {author} {\bibfnamefont {M.}~\bibnamefont
			{Manceau}}, \bibinfo {author} {\bibfnamefont {K.~Y.}\ \bibnamefont
			{Spasibko}}, \bibinfo {author} {\bibfnamefont {G.}~\bibnamefont {Leuchs}},
		\bibinfo {author} {\bibfnamefont {R.}~\bibnamefont {Filip}},\ and\ \bibinfo
		{author} {\bibfnamefont {M.~V.}\ \bibnamefont {Chekhova}},\ }\bibfield
	{title} {\bibinfo {title} {Indefinite-{Mean} {Pareto} {Photon} {Distribution}
			from {Amplified} {Quantum} {Noise}},\ }\href
	{https://doi.org/10.1103/PhysRevLett.123.123606} {\bibfield  {journal}
		{\bibinfo  {journal} {Physical Review Letters}\ }\textbf {\bibinfo {volume}
			{123}},\ \bibinfo {pages} {123606} (\bibinfo {year} {2019})}\BibitemShut
	{NoStop}%
	\bibitem [{\citenamefont {Sharapova}\ \emph {et~al.}(2020)\citenamefont
		{Sharapova}, \citenamefont {Frascella}, \citenamefont {Riabinin},
		\citenamefont {P{\'e}rez}, \citenamefont {Tikhonova}, \citenamefont
		{Lemieux}, \citenamefont {Boyd}, \citenamefont {Leuchs},\ and\ \citenamefont
		{Chekhova}}]{sharapova2020properties}%
	\BibitemOpen
	\bibfield  {author} {\bibinfo {author} {\bibfnamefont {P.~R.}\ \bibnamefont
			{Sharapova}}, \bibinfo {author} {\bibfnamefont {G.}~\bibnamefont
			{Frascella}}, \bibinfo {author} {\bibfnamefont {M.}~\bibnamefont {Riabinin}},
		\bibinfo {author} {\bibfnamefont {A.~M.}\ \bibnamefont {P{\'e}rez}}, \bibinfo
		{author} {\bibfnamefont {O.~V.}\ \bibnamefont {Tikhonova}}, \bibinfo {author}
		{\bibfnamefont {S.}~\bibnamefont {Lemieux}}, \bibinfo {author} {\bibfnamefont
			{R.~W.}\ \bibnamefont {Boyd}}, \bibinfo {author} {\bibfnamefont
			{G.}~\bibnamefont {Leuchs}},\ and\ \bibinfo {author} {\bibfnamefont {M.~V.}\
			\bibnamefont {Chekhova}},\ }\bibfield  {title} {\bibinfo {title} {Properties
			of bright squeezed vacuum at increasing brightness},\ }\href
	{https://link.aps.org/doi/10.1103/PhysRevResearch.2.013371} {\bibfield
		{journal} {\bibinfo  {journal} {Physical Review Research}\ }\textbf {\bibinfo
			{volume} {2}},\ \bibinfo {pages} {013371} (\bibinfo {year}
		{2020})}\BibitemShut {NoStop}%
	\bibitem [{\citenamefont {Rasputnyi}\ \emph {et~al.}(2024)\citenamefont
		{Rasputnyi}, \citenamefont {Chen}, \citenamefont {Birk}, \citenamefont
		{Cohen}, \citenamefont {Kaminer}, \citenamefont {Kr{\"u}ger}, \citenamefont
		{Seletskiy}, \citenamefont {Chekhova},\ and\ \citenamefont
		{Tani}}]{rasputnyi2024high}%
	\BibitemOpen
	\bibfield  {author} {\bibinfo {author} {\bibfnamefont {A.}~\bibnamefont
			{Rasputnyi}}, \bibinfo {author} {\bibfnamefont {Z.}~\bibnamefont {Chen}},
		\bibinfo {author} {\bibfnamefont {M.}~\bibnamefont {Birk}}, \bibinfo {author}
		{\bibfnamefont {O.}~\bibnamefont {Cohen}}, \bibinfo {author} {\bibfnamefont
			{I.}~\bibnamefont {Kaminer}}, \bibinfo {author} {\bibfnamefont
			{M.}~\bibnamefont {Kr{\"u}ger}}, \bibinfo {author} {\bibfnamefont
			{D.}~\bibnamefont {Seletskiy}}, \bibinfo {author} {\bibfnamefont
			{M.}~\bibnamefont {Chekhova}},\ and\ \bibinfo {author} {\bibfnamefont
			{F.}~\bibnamefont {Tani}},\ }\bibfield  {title} {\bibinfo {title}
		{High-harmonic generation by a bright squeezed vacuum},\ }\href
	{https://www.nature.com/articles/s41567-024-02659-x} {\bibfield  {journal}
		{\bibinfo  {journal} {Nature Physics}\ }\textbf {\bibinfo {volume} {20}},\
		\bibinfo {pages} {1960} (\bibinfo {year} {2024})}\BibitemShut {NoStop}%
	\bibitem [{\citenamefont {Lyu}\ \emph {et~al.}(2026)\citenamefont {Lyu},
		\citenamefont {Sun}, \citenamefont {Yi}, \citenamefont {Li}, \citenamefont
		{Liu}, \citenamefont {He}, \citenamefont {Gong}, \citenamefont {Ivanov},\
		and\ \citenamefont {Liu}}]{lyu2026attosecond}%
	\BibitemOpen
	\bibfield  {author} {\bibinfo {author} {\bibfnamefont {Z.}~\bibnamefont
			{Lyu}}, \bibinfo {author} {\bibfnamefont {F.}~\bibnamefont {Sun}}, \bibinfo
		{author} {\bibfnamefont {S.}~\bibnamefont {Yi}}, \bibinfo {author}
		{\bibfnamefont {J.}~\bibnamefont {Li}}, \bibinfo {author} {\bibfnamefont
			{H.}~\bibnamefont {Liu}}, \bibinfo {author} {\bibfnamefont {Q.}~\bibnamefont
			{He}}, \bibinfo {author} {\bibfnamefont {Q.}~\bibnamefont {Gong}}, \bibinfo
		{author} {\bibfnamefont {M.}~\bibnamefont {Ivanov}},\ and\ \bibinfo {author}
		{\bibfnamefont {Y.}~\bibnamefont {Liu}},\ }\href
	{http://arxiv.org/abs/2604.06707} {\bibinfo {title} {Attosecond quantum
			spectroscopy with entangled photon pairs}} (\bibinfo {year} {2026}),\ \Eprint
	{https://arxiv.org/abs/2604.06707} {arXiv:2604.06707 [physics]} \BibitemShut
	{NoStop}%
	\bibitem [{\citenamefont {Lemieux}\ \emph {et~al.}(2025)\citenamefont
		{Lemieux}, \citenamefont {Jalil}, \citenamefont {Purschke}, \citenamefont
		{Boroumand}, \citenamefont {Hammond}, \citenamefont {Villeneuve},
		\citenamefont {Naumov}, \citenamefont {Brabec},\ and\ \citenamefont
		{Vampa}}]{lemieux2025photon}%
	\BibitemOpen
	\bibfield  {author} {\bibinfo {author} {\bibfnamefont {S.}~\bibnamefont
			{Lemieux}}, \bibinfo {author} {\bibfnamefont {S.~A.}\ \bibnamefont {Jalil}},
		\bibinfo {author} {\bibfnamefont {D.~N.}\ \bibnamefont {Purschke}}, \bibinfo
		{author} {\bibfnamefont {N.}~\bibnamefont {Boroumand}}, \bibinfo {author}
		{\bibfnamefont {T.~J.}\ \bibnamefont {Hammond}}, \bibinfo {author}
		{\bibfnamefont {D.}~\bibnamefont {Villeneuve}}, \bibinfo {author}
		{\bibfnamefont {A.}~\bibnamefont {Naumov}}, \bibinfo {author} {\bibfnamefont
			{T.}~\bibnamefont {Brabec}},\ and\ \bibinfo {author} {\bibfnamefont
			{G.}~\bibnamefont {Vampa}},\ }\bibfield  {title} {\bibinfo {title} {Photon
			bunching in high-harmonic emission controlled by quantum light},\ }\href
	{https://doi.org/10.1038/s41566-025-01673-6} {\bibfield  {journal} {\bibinfo
			{journal} {Nature Photonics}\ }\textbf {\bibinfo {volume} {19}},\ \bibinfo
		{pages} {767} (\bibinfo {year} {2025})}\BibitemShut {NoStop}%
	\bibitem [{\citenamefont {Heimerl}\ \emph {et~al.}(2025)\citenamefont
		{Heimerl}, \citenamefont {Rasputnyi}, \citenamefont {P{\"o}lloth},
		\citenamefont {Meier}, \citenamefont {Chekhova},\ and\ \citenamefont
		{Hommelhoff}}]{heimerl2025quantum}%
	\BibitemOpen
	\bibfield  {author} {\bibinfo {author} {\bibfnamefont {J.}~\bibnamefont
			{Heimerl}}, \bibinfo {author} {\bibfnamefont {A.}~\bibnamefont {Rasputnyi}},
		\bibinfo {author} {\bibfnamefont {J.}~\bibnamefont {P{\"o}lloth}}, \bibinfo
		{author} {\bibfnamefont {S.}~\bibnamefont {Meier}}, \bibinfo {author}
		{\bibfnamefont {M.}~\bibnamefont {Chekhova}},\ and\ \bibinfo {author}
		{\bibfnamefont {P.}~\bibnamefont {Hommelhoff}},\ }\bibfield  {title}
	{\bibinfo {title} {Quantum light drives electrons strongly at metal needle
			tips},\ }\href {https://www.nature.com/articles/s41567-025-03087-1}
	{\bibfield  {journal} {\bibinfo  {journal} {Nature Physics}\ } (\bibinfo
		{year} {2025})}\BibitemShut {NoStop}%
	\bibitem [{\citenamefont {Jiang}\ \emph {et~al.}(2026)\citenamefont {Jiang},
		\citenamefont {Pan}, \citenamefont {Chen}, \citenamefont {Zhu}, \citenamefont
		{Zhao}, \citenamefont {Wang}, \citenamefont {Zhang}, \citenamefont {Lu},
		\citenamefont {Han}, \citenamefont {Xiong}, \citenamefont {Wu}, \citenamefont
		{Li}, \citenamefont {Jiang}, \citenamefont {Ni},\ and\ \citenamefont
		{Wu}}]{jiang2026quantumboosted}%
	\BibitemOpen
	\bibfield  {author} {\bibinfo {author} {\bibfnamefont {Z.}~\bibnamefont
			{Jiang}}, \bibinfo {author} {\bibfnamefont {S.}~\bibnamefont {Pan}}, \bibinfo
		{author} {\bibfnamefont {J.}~\bibnamefont {Chen}}, \bibinfo {author}
		{\bibfnamefont {M.}~\bibnamefont {Zhu}}, \bibinfo {author} {\bibfnamefont
			{C.}~\bibnamefont {Zhao}}, \bibinfo {author} {\bibfnamefont {Y.}~\bibnamefont
			{Wang}}, \bibinfo {author} {\bibfnamefont {R.}~\bibnamefont {Zhang}},
		\bibinfo {author} {\bibfnamefont {J.}~\bibnamefont {Lu}}, \bibinfo {author}
		{\bibfnamefont {L.}~\bibnamefont {Han}}, \bibinfo {author} {\bibfnamefont
			{S.}~\bibnamefont {Xiong}}, \bibinfo {author} {\bibfnamefont
			{D.}~\bibnamefont {Wu}}, \bibinfo {author} {\bibfnamefont {W.}~\bibnamefont
			{Li}}, \bibinfo {author} {\bibfnamefont {S.}~\bibnamefont {Jiang}}, \bibinfo
		{author} {\bibfnamefont {H.}~\bibnamefont {Ni}},\ and\ \bibinfo {author}
		{\bibfnamefont {J.}~\bibnamefont {Wu}},\ }\bibfield  {title} {\bibinfo {title}
		{Nonlinear atomic tunnelling boosted by bright squeezed vacuum},\ }\href
	{https://doi.org/10.1038/s41586-026-10485-9} {\bibfield  {journal} {\bibinfo
			{journal} {Nature}\ }(\bibinfo {year} {2026})}\BibitemShut {NoStop}%
	\bibitem [{\citenamefont {Liu}\ \emph {et~al.}(2026)\citenamefont {Liu},
		\citenamefont {Long}, \citenamefont {Li}, \citenamefont {Lyu},\ and\
		\citenamefont {Liu}}]{liu2026strongfield}%
	\BibitemOpen
	\bibfield  {author} {\bibinfo {author} {\bibfnamefont {H.}~\bibnamefont
			{Liu}}, \bibinfo {author} {\bibfnamefont {X.}~\bibnamefont {Long}}, \bibinfo
		{author} {\bibfnamefont {P.}~\bibnamefont {Li}}, \bibinfo {author}
		{\bibfnamefont {Z.}~\bibnamefont {Lyu}},\ and\ \bibinfo {author}
		{\bibfnamefont {Y.}~\bibnamefont {Liu}},\ }\href
	{https://arxiv.org/abs/2604.06703} {\bibinfo {title} {Strong-field ionization
			of atoms with bright squeezed vacuum light}} (\bibinfo {year}
	{2026})\BibitemShut {NoStop}%
	\bibitem [{\citenamefont {Gorlach}\ \emph {et~al.}(2023)\citenamefont
		{Gorlach}, \citenamefont {Tzur}, \citenamefont {Birk}, \citenamefont
		{Kr{\"u}ger}, \citenamefont {Rivera}, \citenamefont {Cohen},\ and\
		\citenamefont {Kaminer}}]{gorlach2023high}%
	\BibitemOpen
	\bibfield  {author} {\bibinfo {author} {\bibfnamefont {A.}~\bibnamefont
			{Gorlach}}, \bibinfo {author} {\bibfnamefont {M.~E.}\ \bibnamefont {Tzur}},
		\bibinfo {author} {\bibfnamefont {M.}~\bibnamefont {Birk}}, \bibinfo {author}
		{\bibfnamefont {M.}~\bibnamefont {Kr{\"u}ger}}, \bibinfo {author}
		{\bibfnamefont {N.}~\bibnamefont {Rivera}}, \bibinfo {author} {\bibfnamefont
			{O.}~\bibnamefont {Cohen}},\ and\ \bibinfo {author} {\bibfnamefont
			{I.}~\bibnamefont {Kaminer}},\ }\bibfield  {title} {\bibinfo {title}
		{High-harmonic generation driven by quantum light},\ }\href
	{https://www.nature.com/articles/s41567-023-02127-y} {\bibfield  {journal}
		{\bibinfo  {journal} {Nature Physics}\ } (\bibinfo {year}
		{2023})}\BibitemShut {NoStop}%
	\bibitem [{\citenamefont {Even~Tzur}\ \emph {et~al.}(2023)\citenamefont
		{Even~Tzur}, \citenamefont {Birk}, \citenamefont {Gorlach}, \citenamefont
		{Kr{\"u}ger}, \citenamefont {Kaminer},\ and\ \citenamefont
		{Cohen}}]{even2023photon}%
	\BibitemOpen
	\bibfield  {author} {\bibinfo {author} {\bibfnamefont {M.}~\bibnamefont
			{Even~Tzur}}, \bibinfo {author} {\bibfnamefont {M.}~\bibnamefont {Birk}},
		\bibinfo {author} {\bibfnamefont {A.}~\bibnamefont {Gorlach}}, \bibinfo
		{author} {\bibfnamefont {M.}~\bibnamefont {Kr{\"u}ger}}, \bibinfo {author}
		{\bibfnamefont {I.}~\bibnamefont {Kaminer}},\ and\ \bibinfo {author}
		{\bibfnamefont {O.}~\bibnamefont {Cohen}},\ }\bibfield  {title} {\bibinfo
		{title} {Photon-statistics force in ultrafast electron dynamics},\ }\href
	{https://www.nature.com/articles/s41566-023-01209-w} {\bibfield  {journal}
		{\bibinfo  {journal} {Nature Photonics}\ }\textbf {\bibinfo {volume} {17}},\
		\bibinfo {pages} {501} (\bibinfo {year} {2023})}\BibitemShut {NoStop}%
	\bibitem [{\citenamefont {Fang}\ \emph {et~al.}(2023)\citenamefont {Fang},
		\citenamefont {Sun}, \citenamefont {He},\ and\ \citenamefont
		{Liu}}]{fang2023strong}%
	\BibitemOpen
	\bibfield  {author} {\bibinfo {author} {\bibfnamefont {Y.}~\bibnamefont
			{Fang}}, \bibinfo {author} {\bibfnamefont {F.-X.}\ \bibnamefont {Sun}},
		\bibinfo {author} {\bibfnamefont {Q.}~\bibnamefont {He}},\ and\ \bibinfo
		{author} {\bibfnamefont {Y.}~\bibnamefont {Liu}},\ }\bibfield  {title}
	{\bibinfo {title} {Strong-{{Field Ionization}} of {{Hydrogen Atoms}} with
			{{Quantum Light}}},\ }\href
	{https://link.aps.org/doi/10.1103/PhysRevLett.130.253201} {\bibfield
		{journal} {\bibinfo  {journal} {Physical Review Letters}\ }\textbf {\bibinfo
			{volume} {130}},\ \bibinfo {pages} {253201} (\bibinfo {year}
		{2023})}\BibitemShut {NoStop}%
	\bibitem [{\citenamefont {Wang}\ and\ \citenamefont
		{Lai}(2023)}]{wang2023high}%
	\BibitemOpen
	\bibfield  {author} {\bibinfo {author} {\bibfnamefont {S.}~\bibnamefont
			{Wang}}\ and\ \bibinfo {author} {\bibfnamefont {X.}~\bibnamefont {Lai}},\
	}\bibfield  {title} {\bibinfo {title} {High-order above-threshold ionization
			of an atom in intense quantum light},\ }\href
	{https://link.aps.org/doi/10.1103/PhysRevA.108.063101} {\bibfield  {journal}
		{\bibinfo  {journal} {Physical Review A}\ }\textbf {\bibinfo {volume}
			{108}},\ \bibinfo {pages} {063101} (\bibinfo {year} {2023})}\BibitemShut
	{NoStop}%
	\bibitem [{\citenamefont {Lyu}\ \emph {et~al.}(2025)\citenamefont {Lyu},
		\citenamefont {Sun}, \citenamefont {Fang}, \citenamefont {He},\ and\
		\citenamefont {Liu}}]{lyu2025effect}%
	\BibitemOpen
	\bibfield  {author} {\bibinfo {author} {\bibfnamefont {Z.}~\bibnamefont
			{Lyu}}, \bibinfo {author} {\bibfnamefont {F.}~\bibnamefont {Sun}}, \bibinfo
		{author} {\bibfnamefont {Y.}~\bibnamefont {Fang}}, \bibinfo {author}
		{\bibfnamefont {Q.}~\bibnamefont {He}},\ and\ \bibinfo {author}
		{\bibfnamefont {Y.}~\bibnamefont {Liu}},\ }\bibfield  {title} {\bibinfo
		{title} {Effect of photon quantum statistics on electrons in above-threshold
			ionization},\ }\href
	{https://link.aps.org/doi/10.1103/PhysRevResearch.7.L012072} {\bibfield
		{journal} {\bibinfo  {journal} {Physical Review Research}\ }\textbf {\bibinfo
			{volume} {7}},\ \bibinfo {pages} {L012072} (\bibinfo {year}
		{2025})}\BibitemShut {NoStop}%
	\bibitem [{\citenamefont {Liu}\ \emph {et~al.}(2025)\citenamefont {Liu},
		\citenamefont {Zhang}, \citenamefont {Wang},\ and\ \citenamefont
		{Yuan}}]{liu2025atomic}%
	\BibitemOpen
	\bibfield  {author} {\bibinfo {author} {\bibfnamefont {H.}~\bibnamefont
			{Liu}}, \bibinfo {author} {\bibfnamefont {H.}~\bibnamefont {Zhang}}, \bibinfo
		{author} {\bibfnamefont {X.}~\bibnamefont {Wang}},\ and\ \bibinfo {author}
		{\bibfnamefont {J.}~\bibnamefont {Yuan}},\ }\bibfield  {title} {\bibinfo
		{title} {Atomic {{Double Ionization}} with {{Quantum Light}}},\ }\href
	{https://link.aps.org/doi/10.1103/PhysRevLett.134.123202} {\bibfield
		{journal} {\bibinfo  {journal} {Physical Review Letters}\ }\textbf {\bibinfo
			{volume} {134}},\ \bibinfo {pages} {123202} (\bibinfo {year}
		{2025})}\BibitemShut {NoStop}%
	\bibitem [{\citenamefont {Li}\ \emph {et~al.}(2014)\citenamefont {Li},
		\citenamefont {Tong}, \citenamefont {Morishita}, \citenamefont {Wei},\ and\
		\citenamefont {Lin}}]{li2014fine}%
	\BibitemOpen
	\bibfield  {author} {\bibinfo {author} {\bibfnamefont {Q.}~\bibnamefont
			{Li}}, \bibinfo {author} {\bibfnamefont {X.-M.}\ \bibnamefont {Tong}},
		\bibinfo {author} {\bibfnamefont {T.}~\bibnamefont {Morishita}}, \bibinfo
		{author} {\bibfnamefont {H.}~\bibnamefont {Wei}},\ and\ \bibinfo {author}
		{\bibfnamefont {C.~D.}\ \bibnamefont {Lin}},\ }\bibfield  {title} {\bibinfo
		{title} {Fine structures in the intensity dependence of excitation and
			ionization probabilities of hydrogen atoms in intense 800-nm laser pulses},\
	}\href {https://link.aps.org/doi/10.1103/PhysRevA.89.023421} {\bibfield
		{journal} {\bibinfo  {journal} {Physical Review A}\ }\textbf {\bibinfo
			{volume} {89}},\ \bibinfo {pages} {023421} (\bibinfo {year}
		{2014})}\BibitemShut {NoStop}%
	\bibitem [{\citenamefont {Chetty}\ \emph {et~al.}(2020)\citenamefont {Chetty},
		\citenamefont {Glover}, \citenamefont {{deHarak}}, \citenamefont {Tong},
		\citenamefont {Xu}, \citenamefont {Pauly}, \citenamefont {Smith},
		\citenamefont {Hamilton}, \citenamefont {Bartschat}, \citenamefont {Ziegel},
		\citenamefont {Douguet}, \citenamefont {Luiten}, \citenamefont {Light},
		\citenamefont {Litvinyuk},\ and\ \citenamefont
		{Sang}}]{chetty2020observation}%
	\BibitemOpen
	\bibfield  {author} {\bibinfo {author} {\bibfnamefont {D.}~\bibnamefont
			{Chetty}}, \bibinfo {author} {\bibfnamefont {R.~D.}\ \bibnamefont {Glover}},
		\bibinfo {author} {\bibfnamefont {B.~A.}\ \bibnamefont {{deHarak}}}, \bibinfo
		{author} {\bibfnamefont {X.~M.}\ \bibnamefont {Tong}}, \bibinfo {author}
		{\bibfnamefont {H.}~\bibnamefont {Xu}}, \bibinfo {author} {\bibfnamefont
			{T.}~\bibnamefont {Pauly}}, \bibinfo {author} {\bibfnamefont
			{N.}~\bibnamefont {Smith}}, \bibinfo {author} {\bibfnamefont {K.~R.}\
			\bibnamefont {Hamilton}}, \bibinfo {author} {\bibfnamefont {K.}~\bibnamefont
			{Bartschat}}, \bibinfo {author} {\bibfnamefont {J.~P.}\ \bibnamefont
			{Ziegel}}, \bibinfo {author} {\bibfnamefont {N.}~\bibnamefont {Douguet}},
		\bibinfo {author} {\bibfnamefont {A.~N.}\ \bibnamefont {Luiten}}, \bibinfo
		{author} {\bibfnamefont {P.~S.}\ \bibnamefont {Light}}, \bibinfo {author}
		{\bibfnamefont {I.~V.}\ \bibnamefont {Litvinyuk}},\ and\ \bibinfo {author}
		{\bibfnamefont {R.~T.}\ \bibnamefont {Sang}},\ }\bibfield  {title} {\bibinfo
		{title} {Observation of dynamic {{Stark}} resonances in strong-field
			excitation},\ }\href {https://link.aps.org/doi/10.1103/PhysRevA.101.053402}
	{\bibfield  {journal} {\bibinfo  {journal} {Physical Review A}\ }\textbf
		{\bibinfo {volume} {101}},\ \bibinfo {pages} {053402} (\bibinfo {year}
		{2020})}\BibitemShut {NoStop}%
	\bibitem [{\citenamefont {Zhang}\ \emph {et~al.}(2014)\citenamefont {Zhang},
		\citenamefont {Chen},\ and\ \citenamefont {Zhao}}]{zhang2014generation}%
	\BibitemOpen
	\bibfield  {author} {\bibinfo {author} {\bibfnamefont {B.}~\bibnamefont
			{Zhang}}, \bibinfo {author} {\bibfnamefont {W.}~\bibnamefont {Chen}},\ and\
		\bibinfo {author} {\bibfnamefont {Z.}~\bibnamefont {Zhao}},\ }\bibfield
	{title} {\bibinfo {title} {Generation of {{Rydberg}} states of hydrogen atoms
			with intense laser pulses: {{The}} roles of {{Coulomb}} force and initial
			lateral momentum},\ }\href
	{https://link.aps.org/doi/10.1103/PhysRevA.90.023409} {\bibfield  {journal}
		{\bibinfo  {journal} {Physical Review A}\ }\textbf {\bibinfo {volume} {90}},\
		\bibinfo {pages} {023409} (\bibinfo {year} {2014})}\BibitemShut {NoStop}%
	\bibitem [{\citenamefont {Hu}\ \emph {et~al.}(2019)\citenamefont {Hu},
		\citenamefont {Hao}, \citenamefont {Lv}, \citenamefont {Liu}, \citenamefont
		{Yang}, \citenamefont {Xu}, \citenamefont {Jin}, \citenamefont {Ding},
		\citenamefont {Li}, \citenamefont {Li} \emph {et~al.}}]{hu2019quantum}%
	\BibitemOpen
	\bibfield  {author} {\bibinfo {author} {\bibfnamefont {S.}~\bibnamefont
			{Hu}}, \bibinfo {author} {\bibfnamefont {X.}~\bibnamefont {Hao}}, \bibinfo
		{author} {\bibfnamefont {H.}~\bibnamefont {Lv}}, \bibinfo {author}
		{\bibfnamefont {M.}~\bibnamefont {Liu}}, \bibinfo {author} {\bibfnamefont
			{T.}~\bibnamefont {Yang}}, \bibinfo {author} {\bibfnamefont {H.}~\bibnamefont
			{Xu}}, \bibinfo {author} {\bibfnamefont {M.}~\bibnamefont {Jin}}, \bibinfo
		{author} {\bibfnamefont {D.}~\bibnamefont {Ding}}, \bibinfo {author}
		{\bibfnamefont {Q.}~\bibnamefont {Li}}, \bibinfo {author} {\bibfnamefont
			{W.}~\bibnamefont {Li}}, \emph {et~al.},\ }\bibfield  {title} {\bibinfo
		{title} {Quantum dynamics of atomic rydberg excitation in strong laser
			fields},\ }\href@noop {} {\bibfield  {journal} {\bibinfo  {journal} {Optics
				express}\ }\textbf {\bibinfo {volume} {27}},\ \bibinfo {pages} {31629}
		(\bibinfo {year} {2019})}\BibitemShut {NoStop}%
	\bibitem [{\citenamefont {Chetty}\ \emph {et~al.}(2022)\citenamefont {Chetty},
		\citenamefont {Glover}, \citenamefont {Tong}, \citenamefont {{deHarak}},
		\citenamefont {Xu}, \citenamefont {Haram}, \citenamefont {Bartschat},
		\citenamefont {Palmer}, \citenamefont {Luiten}, \citenamefont {Light},
		\citenamefont {Litvinyuk},\ and\ \citenamefont {Sang}}]{chetty2022carrier}%
	\BibitemOpen
	\bibfield  {author} {\bibinfo {author} {\bibfnamefont {D.}~\bibnamefont
			{Chetty}}, \bibinfo {author} {\bibfnamefont {R.~D.}\ \bibnamefont {Glover}},
		\bibinfo {author} {\bibfnamefont {X.~M.}\ \bibnamefont {Tong}}, \bibinfo
		{author} {\bibfnamefont {B.~A.}\ \bibnamefont {{deHarak}}}, \bibinfo {author}
		{\bibfnamefont {H.}~\bibnamefont {Xu}}, \bibinfo {author} {\bibfnamefont
			{N.}~\bibnamefont {Haram}}, \bibinfo {author} {\bibfnamefont
			{K.}~\bibnamefont {Bartschat}}, \bibinfo {author} {\bibfnamefont {A.~J.}\
			\bibnamefont {Palmer}}, \bibinfo {author} {\bibfnamefont {A.~N.}\
			\bibnamefont {Luiten}}, \bibinfo {author} {\bibfnamefont {P.~S.}\
			\bibnamefont {Light}}, \bibinfo {author} {\bibfnamefont {I.~V.}\ \bibnamefont
			{Litvinyuk}},\ and\ \bibinfo {author} {\bibfnamefont {R.~T.}\ \bibnamefont
			{Sang}},\ }\bibfield  {title} {\bibinfo {title} {Carrier-{{Envelope
					Phase-Dependent Strong-Field Excitation}}},\ }\href
	{https://link.aps.org/doi/10.1103/PhysRevLett.128.173201} {\bibfield
		{journal} {\bibinfo  {journal} {Physical Review Letters}\ }\textbf {\bibinfo
			{volume} {128}},\ \bibinfo {pages} {173201} (\bibinfo {year}
		{2022})}\BibitemShut {NoStop}%
	\bibitem [{\citenamefont {Scully}\ and\ \citenamefont
		{Zubairy}(1997)}]{scully1997quantum}%
	\BibitemOpen
	\bibfield  {author} {\bibinfo {author} {\bibfnamefont {M.~O.}\ \bibnamefont
			{Scully}}\ and\ \bibinfo {author} {\bibfnamefont {M.~S.}\ \bibnamefont
			{Zubairy}},\ }\href@noop {} {\emph {\bibinfo {title} {Quantum optics}}}\
	(\bibinfo  {publisher} {Cambridge university press},\ \bibinfo {year}
	{1997})\BibitemShut {NoStop}%
	\bibitem [{\citenamefont {Hillery}\ \emph {et~al.}(1984)\citenamefont
		{Hillery}, \citenamefont {O'Connell}, \citenamefont {Scully},\ and\
		\citenamefont {Wigner}}]{hillery1984distribution}%
	\BibitemOpen
	\bibfield  {author} {\bibinfo {author} {\bibfnamefont {M.}~\bibnamefont
			{Hillery}}, \bibinfo {author} {\bibfnamefont {R.~F.}\ \bibnamefont
			{O'Connell}}, \bibinfo {author} {\bibfnamefont {M.~O.}\ \bibnamefont
			{Scully}},\ and\ \bibinfo {author} {\bibfnamefont {E.~P.}\ \bibnamefont
			{Wigner}},\ }\bibfield  {title} {\bibinfo {title} {Distribution functions in
			physics: Fundamentals},\ }\href@noop {} {\bibfield  {journal} {\bibinfo
			{journal} {Physics reports}\ }\textbf {\bibinfo {volume} {106}},\ \bibinfo
		{pages} {121} (\bibinfo {year} {1984})}\BibitemShut {NoStop}%
	\bibitem [{\citenamefont {Wang}\ and\ \citenamefont
		{Bian}(2025)}]{wang2025high}%
	\BibitemOpen
	\bibfield  {author} {\bibinfo {author} {\bibfnamefont {Y.-B.}\ \bibnamefont
			{Wang}}\ and\ \bibinfo {author} {\bibfnamefont {X.-B.}\ \bibnamefont
			{Bian}},\ }\bibfield  {title} {\bibinfo {title} {High-order harmonic
			generation in quantum light by a generalized von {{Neumann}} lattice
			method},\ }\href {https://link.aps.org/doi/10.1103/PhysRevA.111.043111}
	{\bibfield  {journal} {\bibinfo  {journal} {Physical Review A}\ }\textbf
		{\bibinfo {volume} {111}},\ \bibinfo {pages} {043111} (\bibinfo {year}
		{2025})}\BibitemShut {NoStop}%
	\bibitem [{\citenamefont {Kim}\ \emph {et~al.}(1989)\citenamefont {Kim},
		\citenamefont {De~Oliveira},\ and\ \citenamefont
		{Knight}}]{kim1989properties}%
	\BibitemOpen
	\bibfield  {author} {\bibinfo {author} {\bibfnamefont {M.}~\bibnamefont
			{Kim}}, \bibinfo {author} {\bibfnamefont {F.}~\bibnamefont {De~Oliveira}},\
		and\ \bibinfo {author} {\bibfnamefont {P.}~\bibnamefont {Knight}},\
	}\bibfield  {title} {\bibinfo {title} {Properties of squeezed number states
			and squeezed thermal states},\ }\href@noop {} {\bibfield  {journal} {\bibinfo
			{journal} {Physical Review A}\ }\textbf {\bibinfo {volume} {40}},\ \bibinfo
		{pages} {2494} (\bibinfo {year} {1989})}\BibitemShut {NoStop}%
	\bibitem [{\citenamefont {Tong}\ and\ \citenamefont
		{Chu}(1997)}]{tong1997theoretical}%
	\BibitemOpen
	\bibfield  {author} {\bibinfo {author} {\bibfnamefont {X.-M.}\ \bibnamefont
			{Tong}}\ and\ \bibinfo {author} {\bibfnamefont {S.-I.}\ \bibnamefont {Chu}},\
	}\bibfield  {title} {\bibinfo {title} {Theoretical study of multiple
			high-order harmonic generation by intense ultrashort pulsed laser fields: A
			new generalized pseudospectral time-dependent method},\ }\href@noop {}
	{\bibfield  {journal} {\bibinfo  {journal} {Chemical Physics}\ }\textbf
		{\bibinfo {volume} {217}},\ \bibinfo {pages} {119} (\bibinfo {year}
		{1997})}\BibitemShut {NoStop}%
	\bibitem [{\citenamefont {KELDYSH}(1965)}]{keldysh1965ionization}%
	\BibitemOpen
	\bibfield  {author} {\bibinfo {author} {\bibfnamefont {L.}~\bibnamefont
			{KELDYSH}},\ }\bibfield  {title} {\bibinfo {title} {Ionization in the field
			of a strong electromagnetic wave},\ }\href@noop {} {\bibfield  {journal}
		{\bibinfo  {journal} {SOVIET PHYSICS JETP}\ }\textbf {\bibinfo {volume} {20}}
		(\bibinfo {year} {1965})}\BibitemShut {NoStop}%
	\bibitem [{\citenamefont {Bing-Bing}\ \emph {et~al.}(2006)\citenamefont
		{Bing-Bing}, \citenamefont {Xiao-Feng}, \citenamefont {Pan-Ming},
		\citenamefont {Jing},\ and\ \citenamefont {Jie}}]{bing2006coulomb}%
	\BibitemOpen
	\bibfield  {author} {\bibinfo {author} {\bibfnamefont {W.}~\bibnamefont
			{Bing-Bing}}, \bibinfo {author} {\bibfnamefont {L.}~\bibnamefont
			{Xiao-Feng}}, \bibinfo {author} {\bibfnamefont {F.}~\bibnamefont {Pan-Ming}},
		\bibinfo {author} {\bibfnamefont {C.}~\bibnamefont {Jing}},\ and\ \bibinfo
		{author} {\bibfnamefont {L.}~\bibnamefont {Jie}},\ }\bibfield  {title}
	{\bibinfo {title} {Coulomb potential recapture effect in above-barrier
			ionizationin laser pulses},\ }\href@noop {} {\bibfield  {journal} {\bibinfo
			{journal} {Chinese Physics Letters}\ }\textbf {\bibinfo {volume} {23}},\
		\bibinfo {pages} {2729} (\bibinfo {year} {2006})}\BibitemShut {NoStop}%
	\bibitem [{\citenamefont {Nubbemeyer}\ \emph {et~al.}(2008)\citenamefont
		{Nubbemeyer}, \citenamefont {Gorling}, \citenamefont {Saenz}, \citenamefont
		{Eichmann},\ and\ \citenamefont {Sandner}}]{nubbemeyer2008strong}%
	\BibitemOpen
	\bibfield  {author} {\bibinfo {author} {\bibfnamefont {T.}~\bibnamefont
			{Nubbemeyer}}, \bibinfo {author} {\bibfnamefont {K.}~\bibnamefont {Gorling}},
		\bibinfo {author} {\bibfnamefont {A.}~\bibnamefont {Saenz}}, \bibinfo
		{author} {\bibfnamefont {U.}~\bibnamefont {Eichmann}},\ and\ \bibinfo
		{author} {\bibfnamefont {W.}~\bibnamefont {Sandner}},\ }\bibfield  {title}
	{\bibinfo {title} {Strong-{{Field Tunneling}} without {{Ionization}}},\
	}\href {https://link.aps.org/doi/10.1103/PhysRevLett.101.233001} {\bibfield
		{journal} {\bibinfo  {journal} {Physical Review Letters}\ }\textbf {\bibinfo
			{volume} {101}},\ \bibinfo {pages} {233001} (\bibinfo {year}
		{2008})}\BibitemShut {NoStop}%
	\bibitem [{\citenamefont {Liu}\ \emph {et~al.}(2021)\citenamefont {Liu},
		\citenamefont {Xu}, \citenamefont {Hu}, \citenamefont {Becker}, \citenamefont
		{Quan}, \citenamefont {Liu},\ and\ \citenamefont {Chen}}]{liu2021electron}%
	\BibitemOpen
	\bibfield  {author} {\bibinfo {author} {\bibfnamefont {M.}~\bibnamefont
			{Liu}}, \bibinfo {author} {\bibfnamefont {S.}~\bibnamefont {Xu}}, \bibinfo
		{author} {\bibfnamefont {S.}~\bibnamefont {Hu}}, \bibinfo {author}
		{\bibfnamefont {W.}~\bibnamefont {Becker}}, \bibinfo {author} {\bibfnamefont
			{W.}~\bibnamefont {Quan}}, \bibinfo {author} {\bibfnamefont {X.}~\bibnamefont
			{Liu}},\ and\ \bibinfo {author} {\bibfnamefont {J.}~\bibnamefont {Chen}},\
	}\bibfield  {title} {\bibinfo {title} {Electron dynamics in laser-driven
			atoms near the continuum threshold},\ }\href
	{https://opg.optica.org/abstract.cfm?URI=optica-8-6-765} {\bibfield
		{journal} {\bibinfo  {journal} {Optica}\ }\textbf {\bibinfo {volume} {8}},\
		\bibinfo {pages} {765} (\bibinfo {year} {2021})}\BibitemShut {NoStop}%
	\bibitem [{\citenamefont {Yi}\ \emph {et~al.}(2026)\citenamefont {Yi}
		\emph {et~al.}}]{yi2026intensity}%
	\BibitemOpen
	\bibfield  {author} {\bibinfo {author} {\bibfnamefont {X.}~\bibnamefont
			{Yi}} \emph {et~al.},\ }\bibfield  {title} {\bibinfo {title}
		{Intensity-dependent interferences in strong-field Rydberg-state excitation},\
	}\href {https://doi.org/10.1103/4d6p-55n1} {\bibfield  {journal} {\bibinfo
			{journal} {Physical Review Research}\ }(\bibinfo {year}
		{2026})}\BibitemShut {NoStop}%
	\bibitem [{\citenamefont {Zimmermann}\ \emph
		{et~al.}(2017)\citenamefont {Zimmermann}, \citenamefont
		{Patchkovskii}, \citenamefont {Ivanov},\ and\ \citenamefont
		{Eichmann}}]{zimmermann_unified_2017}%
	\BibitemOpen
	\bibfield  {author} {\bibinfo {author} {\bibfnamefont {H.}~\bibnamefont
			{Zimmermann}}, \bibinfo {author} {\bibfnamefont {S.}~\bibnamefont
			{Patchkovskii}}, \bibinfo {author} {\bibfnamefont {M.}~\bibnamefont
			{Ivanov}},\ and\ \bibinfo {author} {\bibfnamefont {U.}~\bibnamefont
			{Eichmann}},\ }\bibfield  {title} {\bibinfo {title} {Unified {Time} and
			{Frequency} {Picture} of {Ultrafast} {Atomic} {Excitation} in {Strong}
			{Laser} {Fields}},\ }\href {https://doi.org/10.1103/PhysRevLett.118.013003}
	{\bibfield  {journal} {\bibinfo  {journal} {Physical Review Letters}\
		}\textbf {\bibinfo {volume} {118}},\ \bibinfo {pages} {013003} (\bibinfo
		{year} {2017})}\BibitemShut {NoStop}%
	\bibitem [{\citenamefont {Xu}\ \emph {et~al.}(2020)\citenamefont {Xu},
		\citenamefont {Liu}, \citenamefont {Hu}, \citenamefont {Shu}, \citenamefont
		{Quan}, \citenamefont {Xiao}, \citenamefont {Zhou}, \citenamefont {Wei},
		\citenamefont {Zhao}, \citenamefont {Sun}, \citenamefont {Wang},
		\citenamefont {Hua}, \citenamefont {Gong}, \citenamefont {Lai}, \citenamefont
		{Chen},\ and\ \citenamefont {Liu}}]{xu2020observation}%
	\BibitemOpen
	\bibfield  {author} {\bibinfo {author} {\bibfnamefont {S.}~\bibnamefont
			{Xu}}, \bibinfo {author} {\bibfnamefont {M.}~\bibnamefont {Liu}}, \bibinfo
		{author} {\bibfnamefont {S.}~\bibnamefont {Hu}}, \bibinfo {author}
		{\bibfnamefont {Z.}~\bibnamefont {Shu}}, \bibinfo {author} {\bibfnamefont
			{W.}~\bibnamefont {Quan}}, \bibinfo {author} {\bibfnamefont {Z.}~\bibnamefont
			{Xiao}}, \bibinfo {author} {\bibfnamefont {Y.}~\bibnamefont {Zhou}}, \bibinfo
		{author} {\bibfnamefont {M.}~\bibnamefont {Wei}}, \bibinfo {author}
		{\bibfnamefont {M.}~\bibnamefont {Zhao}}, \bibinfo {author} {\bibfnamefont
			{R.}~\bibnamefont {Sun}}, \bibinfo {author} {\bibfnamefont {Y.}~\bibnamefont
			{Wang}}, \bibinfo {author} {\bibfnamefont {L.}~\bibnamefont {Hua}}, \bibinfo
		{author} {\bibfnamefont {C.}~\bibnamefont {Gong}}, \bibinfo {author}
		{\bibfnamefont {X.}~\bibnamefont {Lai}}, \bibinfo {author} {\bibfnamefont
			{J.}~\bibnamefont {Chen}},\ and\ \bibinfo {author} {\bibfnamefont
			{X.}~\bibnamefont {Liu}},\ }\bibfield  {title} {\bibinfo {title} {Observation
			of a transition in the dynamics of strong-field atomic excitation},\ }\href
	{https://link.aps.org/doi/10.1103/PhysRevA.102.043104} {\bibfield  {journal}
		{\bibinfo  {journal} {Physical Review A}\ }\textbf {\bibinfo {volume}
			{102}},\ \bibinfo {pages} {043104} (\bibinfo {year} {2020})}\BibitemShut
	{NoStop}%
	\bibitem [{\citenamefont {Mollow}(1968)}]{mollow1968two}%
	\BibitemOpen
	\bibfield  {author} {\bibinfo {author} {\bibfnamefont {B.}~\bibnamefont
			{Mollow}},\ }\bibfield  {title} {\bibinfo {title} {Two-photon absorption and
			field correlation functions},\ }\href@noop {} {\bibfield  {journal} {\bibinfo
			{journal} {Physical Review}\ }\textbf {\bibinfo {volume} {175}},\ \bibinfo
		{pages} {1555} (\bibinfo {year} {1968})}\BibitemShut {NoStop}%
	\bibitem [{\citenamefont {Agarwal}(1970)}]{agarwal1970field}%
	\BibitemOpen
	\bibfield  {author} {\bibinfo {author} {\bibfnamefont {G.}~\bibnamefont
			{Agarwal}},\ }\bibfield  {title} {\bibinfo {title} {Field-correlation effects
			in multiphoton absorption processes},\ }\href@noop {} {\bibfield  {journal}
		{\bibinfo  {journal} {Physical Review A}\ }\textbf {\bibinfo {volume} {1}},\
		\bibinfo {pages} {1445} (\bibinfo {year} {1970})}\BibitemShut {NoStop}%
	\bibitem [{\citenamefont {Spasibko}\ \emph {et~al.}(2017)\citenamefont
		{Spasibko}, \citenamefont {Kopylov}, \citenamefont {Krutyanskiy},
		\citenamefont {Murzina}, \citenamefont {Leuchs},\ and\ \citenamefont
		{Chekhova}}]{spasibko_multiphoton_2017}%
	\BibitemOpen
	\bibfield  {author} {\bibinfo {author} {\bibfnamefont {K.~Y.}\ \bibnamefont
			{Spasibko}}, \bibinfo {author} {\bibfnamefont {D.~A.}\ \bibnamefont
			{Kopylov}}, \bibinfo {author} {\bibfnamefont {V.~L.}\ \bibnamefont
			{Krutyanskiy}}, \bibinfo {author} {\bibfnamefont {T.~V.}\ \bibnamefont
			{Murzina}}, \bibinfo {author} {\bibfnamefont {G.}~\bibnamefont {Leuchs}},\
		and\ \bibinfo {author} {\bibfnamefont {M.~V.}\ \bibnamefont {Chekhova}},\
	}\bibfield  {title} {\bibinfo {title} {Multiphoton {Effects} {Enhanced} due
			to {Ultrafast} {Photon}-{Number} {Fluctuations}},\ }\href
	{https://doi.org/10.1103/PhysRevLett.119.223603} {\bibfield  {journal}
		{\bibinfo  {journal} {Physical Review Letters}\ }\textbf {\bibinfo {volume}
			{119}},\ \bibinfo {pages} {223603} (\bibinfo {year} {2017})}\BibitemShut
	{NoStop}%
	\bibitem [{\citenamefont {Zhao}\ \emph {et~al.}(2024)\citenamefont {Zhao},
		\citenamefont {Liu}, \citenamefont {Wang},\ and\ \citenamefont
		{Zhao}}]{zhao2024twin}%
	\BibitemOpen
	\bibfield  {author} {\bibinfo {author} {\bibfnamefont {J.}~\bibnamefont
			{Zhao}}, \bibinfo {author} {\bibfnamefont {J.}~\bibnamefont {Liu}}, \bibinfo
		{author} {\bibfnamefont {X.}~\bibnamefont {Wang}},\ and\ \bibinfo {author}
		{\bibfnamefont {Z.}~\bibnamefont {Zhao}},\ }\bibfield  {title} {\bibinfo
		{title} {Twin-{{Capture Rydberg State Excitation Enhanced}} with {{Few-Cycle
					Laser Pulses}}},\ }\href
	{https://iopscience.iop.org/article/10.1088/0256-307X/41/1/013201} {\bibfield
		{journal} {\bibinfo  {journal} {Chinese Physics Letters}\ }\textbf {\bibinfo
		{volume} {41}},\ \bibinfo {pages} {013201} (\bibinfo {year}
		{2024})}\BibitemShut {NoStop}%
	\bibitem{supplemental_material}%
	See Supplemental Material at [URL will be inserted by publisher] for a
	detailed derivation of the quantum-statistical recapture-interference model.
	\bibitem [{\citenamefont {Arb{\'o}}\ \emph {et~al.}(2008)\citenamefont
		{Arb{\'o}}, \citenamefont {Dimitriou}, \citenamefont {Persson},\ and\
		\citenamefont {Burgd{\"o}rfer}}]{arbo2008sub}%
	\BibitemOpen
	\bibfield  {author} {\bibinfo {author} {\bibfnamefont {D.~G.}\ \bibnamefont
			{Arb{\'o}}}, \bibinfo {author} {\bibfnamefont {K.~I.}\ \bibnamefont
			{Dimitriou}}, \bibinfo {author} {\bibfnamefont {E.}~\bibnamefont {Persson}},\
		and\ \bibinfo {author} {\bibfnamefont {J.}~\bibnamefont {Burgd{\"o}rfer}},\
	}\bibfield  {title} {\bibinfo {title} {Sub-{{Poissonian}} angular momentum
			distribution near threshold in atomic ionization by short laser pulses},\
	}\href {https://link.aps.org/doi/10.1103/PhysRevA.78.013406} {\bibfield
		{journal} {\bibinfo  {journal} {Physical Review A}\ }\textbf {\bibinfo
			{volume} {78}},\ \bibinfo {pages} {013406} (\bibinfo {year}
		{2008})}\BibitemShut {NoStop}%
	\bibitem [{\citenamefont {Venzke}\ \emph {et~al.}(2018)\citenamefont {Venzke},
		\citenamefont {Reiff}, \citenamefont {Xue}, \citenamefont
		{{Jaro{\'n}-Becker}},\ and\ \citenamefont {Becker}}]{venzke2018angular}%
	\BibitemOpen
	\bibfield  {author} {\bibinfo {author} {\bibfnamefont {J.}~\bibnamefont
			{Venzke}}, \bibinfo {author} {\bibfnamefont {R.}~\bibnamefont {Reiff}},
		\bibinfo {author} {\bibfnamefont {Z.}~\bibnamefont {Xue}}, \bibinfo {author}
		{\bibfnamefont {A.}~\bibnamefont {{Jaro{\'n}-Becker}}},\ and\ \bibinfo
		{author} {\bibfnamefont {A.}~\bibnamefont {Becker}},\ }\bibfield  {title}
	{\bibinfo {title} {Angular momentum distribution in {{Rydberg}} states
			excited by a strong laser pulse},\ }\href
	{https://link.aps.org/doi/10.1103/PhysRevA.98.043434} {\bibfield  {journal}
		{\bibinfo  {journal} {Physical Review A}\ }\textbf {\bibinfo {volume} {98}},\
		\bibinfo {pages} {043434} (\bibinfo {year} {2018})}\BibitemShut {NoStop}%
	\bibitem [{\citenamefont {Liu}\ \emph {et~al.}(2024)\citenamefont {Liu},
		\citenamefont {Yi}, \citenamefont {Chen}, \citenamefont {Sun}, \citenamefont
		{Lv}, \citenamefont {Hu}, \citenamefont {Becker}, \citenamefont {Xu},\ and\
		\citenamefont {Chen}}]{liu2024parity}%
	\BibitemOpen
	\bibfield  {author} {\bibinfo {author} {\bibfnamefont {Y.}~\bibnamefont
			{Liu}}, \bibinfo {author} {\bibfnamefont {X.}~\bibnamefont {Yi}}, \bibinfo
		{author} {\bibfnamefont {Q.}~\bibnamefont {Chen}}, \bibinfo {author}
		{\bibfnamefont {T.}~\bibnamefont {Sun}}, \bibinfo {author} {\bibfnamefont
			{H.}~\bibnamefont {Lv}}, \bibinfo {author} {\bibfnamefont {S.}~\bibnamefont
			{Hu}}, \bibinfo {author} {\bibfnamefont {W.}~\bibnamefont {Becker}}, \bibinfo
		{author} {\bibfnamefont {H.}~\bibnamefont {Xu}},\ and\ \bibinfo {author}
		{\bibfnamefont {J.}~\bibnamefont {Chen}},\ }\bibfield  {title} {\bibinfo
		{title} {Parity effects in {{Rydberg-state}} excitation in intense laser
			fields},\ }\href {https://opg.optica.org/abstract.cfm?URI=prj-12-12-3033}
	{\bibfield  {journal} {\bibinfo  {journal} {Photonics Research}\ }\textbf
		{\bibinfo {volume} {12}},\ \bibinfo {pages} {3033} (\bibinfo {year}
		{2024})}\BibitemShut {NoStop}%
\end{thebibliography}
%

\end{document}